\documentclass[amsmath,amssymb,aps,twocolumn,notitlepage,longbibliography,
showpacs,pre,floatfix]{revtex4-2}

\usepackage[T1]{fontenc}

\usepackage{graphicx}
\usepackage[export]{adjustbox}
\usepackage{caption}
\usepackage{subfig}
\usepackage{dcolumn}
\usepackage{bm}
\usepackage{xcolor}
\usepackage{subfig}
\usepackage{lipsum}
\usepackage[normalem]{ulem}

\begin{document}
\preprint{APS/123-QED}
\author{J. Talbot}
\affiliation{Laboratoire de Physique Th\'eorique de la Mati\`ere Condens\'ee (UMR CNRS 7600), Sorbonne Universit\'e, 4 Place Jussieu, 75252 Paris Cedex 05, France}
\author{C. Antoine}
\affiliation{Laboratoire de Physique Th\'eorique de la Mati\`ere Condens\'ee (UMR CNRS 7600), Sorbonne Universit\'e, 4 Place Jussieu, 75252 Paris Cedex 05, France}
\author{P. Claudin}
\affiliation{Physique et M\'ecanique des Milieux H\'et\'erog\`enes, PMMH UMR 7636 CNRS, ESPCI Paris, PSL Research University, Sorbonne Universit\'e, Universit\'e Paris Cit\'e, 7 quai St Bernard, 75005 Paris, France}
\author{E. Somfai}
\affiliation{Institute for Solid State Physics and Optics, HUN-REN Wigner Research Centre for Physics, P.O. Box 49, H-1525 Budapest, Hungary}
\author{T. B\"orzs\"onyi}
\affiliation{Institute for Solid State Physics and Optics, HUN-REN Wigner Research Centre for Physics, P.O. Box 49, H-1525 Budapest, Hungary}
% \title{A combined Fokker-Planck and Langevin analysis of noisy Jeffery orbits\\ in 2D and 3D}
\title{Exploring noisy Jeffery orbits: \\A combined Fokker-Planck and Langevin analysis in 2D and 3D}
\date{\today}
\begin{abstract}
The behavior of non-spherical particles in a shear-flow is of significant practical and theoretical interest.  These systems have been the object of numerous investigations since the pioneering work of Jeffery a century ago. His eponymous orbits describe the deterministic motion of an isolated, rod-like particle  in a shear flow. Subsequently, the effect of adding noise was investigated. The theory has been applied to colloidal particles, macromolecules, anisometric granular particles and most recently to microswimmers, for example bacteria.  We study the Jeffery orbits of elongated (uniaxial, prolate) particles subject to noise using Langevin simulations and a Fokker-Planck equation. We extend the analytical solution for infinitely thin needles ($\beta=1$) obtained by Doi and Edwards to particles with arbitrary shape factor ($0\le \beta\le 1$) and validate the theory by comparing it with simulations. We examine  the rotation of the particle around the vorticity axis and study the orientational order matrix. We use the latter to obtain scalar order parameters $s$ and $r$ describing nematic ordering and biaxiality from the orientational distribution function. The value of $s$ (nematic ordering) increases monotonically with increasing P\'eclet number, while $r$ (measure of biaxiality) displays a maximum value. From perturbation theory we obtain simple expressions that provide accurate descriptions at low noise (or large P\'eclet numbers). We also examine the orientational distribution in the v-grad v plane and in the perpendicular direction. 
Finally we present the solution of the Fokker-Planck equation for a strictly two-dimensional (2D) system. For the same noise amplitude the average rotation speed of the particle in 3D is larger than in 2D. 
\end{abstract}
%\begin{document}

%\begin{document}

\maketitle
\section{Introduction}

% The behavior of elongated rods in a shear flow is rather intricate and is considered as a fundamental problem in fluid dynamics. Numerous articles presenting theoretical and experimental studies have been published over the years and last year marked the hundredth anniversary of Jeffery's landmark paper in which he presented a mathematical description of the motion of an ellipsoid of revolution in a shear flow. Aside from their fundamental interest, understanding the dynamics of such systems has implications for a wide range of applications including rheology, self-assembly and active particles

Two years ago marked the one hundredth anniversary of G. B. Jeffery's landmark paper \cite{jeffery1922motion}, "The motion of ellipsoidal particles immersed in a viscous fluid". By calculating the torque on the particle in an unbounded fluid subject to a linear shear flow (creeping flow, or zero Reynold's number), he obtained a deterministic description of its time-dependent orientation.  The particle's axis of revolution executes one of an infinite, one parameter, family of  periodic orbits characterized by an orbital constant. The distribution of orientations is anisotropic, i.e certain orientations are preferred over others and this anisotropy increases with increasing elongation. There is no orbit for which the particle ceases to rotate. 

%Philippe:
%Even a century after its publication, Jeffery’s work is still relevant to many current research fields for a deep understanding of the behaviour of sheared elongated particles, including rheology of granular flows and suspensions [cite first block of refs], nanoparticles, e.g. silica micro-rods, in channels [15–17], red blood cells [18], as well as self-assembly and systems of active particles such as microswimmers [cite 19-23 + second block of refs]. The effect of inertia ...

%Julian:
%Even a century after its publication, Jeffery's work is still relevant to many current research fields including nanoparticles, e.g. silica micro-rods,  in channels cite{zottl2019dynamics,Tohme2021transport,einarsson2016tumbling}, and red blood cells \cite{cordasco2016dynamics}. Recently several articles have modelled the motion of microswimmers using Jeffery orbits with noise \cite{junot2019swimming,kaya2009characterization,ganesh2023numerical,berman2022swimmer,guzman2019nonideal}. The effect of inertia \cite{yu2007rotation,lundell2010heavy} and shear-thinning fluids \cite{abtahi2019jeffery}. 

%Merged
Even a century after its publication, Jeffery’s work is still relevant to many current research fields for a deep understanding of the behavior of sheared elongated particles, including rheology of granular flows and suspensions \cite{borzsonyiPRE2012,azemaPRE2012,botonPRE2013,tapia2JFM017,nagyPRE2017,guazzelliJFM2018,butlerARFM2018,trulssonJFM2018,bounouaJR2019,nagyNJP2020,marschallPRE2019,marschallPRE2020,khanPRF2023}, nanoparticles, e.g. silica micro-rods in channels \cite{zottl2019dynamics,Tohme2021transport,einarsson2016tumbling}, 
red blood cells \cite{cordasco2016dynamics}, as well as self-assembly and systems of active particles such as microswimmers \cite{junot2019swimming,kaya2009characterization,ganesh2023numerical,berman2022swimmer,guzman2019nonideal,saintillan2010extensional,berman2022swimmer,guzman2019nonideal,zottlEPJE2013,ishimotoJPSJ2023,ventrellaCM2023,jing2020chirality}.

Many subsequent articles built on Jeffery's pioneering work. Bretherton \cite{bretherton1962motion} extended the theory to arbitrary particles of revolution and later, triaxial ellipsoids \cite{yarin1997chaotic,lundell2011effect,almondo2018intrinsic}, charged fibers \cite{chen1996rheology}, curved particles \cite{crowdy2016flipping} and time-dependent orientational distributions \cite{leahy2015effect} have been studied. 
Other work focused on how the effect of inertia \cite{yu2007rotation,lundell2010heavy} or the shear-thinning character of the fluid \cite{abtahi2019jeffery} changes the nature of particle rotation.
Also, an alternative derivation of the original equations has been given \cite{junk2007new}.

Of particular interest in the context of the present article are 
the efforts to understand the effect of perturbations induced by fluctuations of the surrounding fluid or interaction with other particles on the trajectories.  In this situation, the particle no longer remains on a single Jeffery orbit. The addition of a rotary Brownian motion therefore leads to steady state which no longer depends on the initial orientation.  Early contributions from Boeder \cite{boeder1932stromungsdoppelbrechung}, Peterlin \cite{peterlin1938viskositat} and Burgers \cite{burgers1938motion} were the first to consider the effect of Brownian motion on Jeffery orbits. Years later Leal and Hinch \cite{hinch1972TheEffect,leal1971effectA} derived approximate expressions for the steady state distribution of orbital constants for weak and intermediate Brownian motion and  applied the results to examine the rheological properties of suspensions of non-spherical particles. 

In a landmark study, Doi and Edwards \cite{doi1978dynamics} developed a theory of rod-like macromolecules in concentrated solution.  Although they do not reference Jeffery's work, a limiting case of their model corresponds to an infinitely thin rod  ($\beta=1$) executing a Jeffery orbit perturbed by Brownian motion. They obtained an analytical solution of the Fokker-Planck equation by expressing the associated operator in terms of the angular momentum operators of quantum mechanics.

While most of the studies concern three-dimensional systems, there is also some interest in 2D shear flows \cite{reddy2009orientational}.
Recently, Marschall et al. \cite{marschallPRE2020} examined the effect of noise on the orientational ordering of isolated discorectangles in a 2D shear flow by performing numerical simulations of the Langevin equation.  
In the absence of noise the peak of the orientational distribution corresponds to particles with their long axis parallel to the flow direction, i.e. when they rotate the slowest. Adding noise results in a widening of the orientational distribution  (decreasing order) and shifts the peak backwards (i.e. before the slowest rotation) \cite{marschallPRE2020,dhont2006rod}. This leads to a larger time averaged rotation speed of the particles.
Interestingly, for a dilute suspension without added noise, increasing the concentration of particles, i.e. increasing the intensity of interaction between neighbors, does not lead to exactly the same effects. While increasing particle concentration leads to a similar shift and widening of the peak of the orientation distribution as for the case of added noise, the order parameter and the average rotation speed change in the opposite way: the increasing influence of the neighbors leads to stronger ordering and to a smaller value of the average rotation speed \cite{marschallPRE2020,tegzeSM2020}.

In dry dense granular flows the intensive particle-particle interactions between neighbors also lead to noisy rotation of the grains. Numerical simulations \cite{campbellPOF2011,reddy2009orientational} and experiments \cite{borzsonyiPRL2012,borzsonyiPRE2012,polNJP2022} showed that both the average orientation angle and the order parameter are relatively independent of the shear rate (inertial number), but they change with particle elongation. For longer particles the order parameter is larger and the average orientation of the particles is closer to the flow direction. The angle dependence of the average rotation speed becomes stronger with increasing particle elongation \cite{borzsonyiPRE2012}.

In this contribution we present several new theoretical and numerical results. In three dimensions, we extend the analytical calculations of Doi and Edwards to particles with arbitrary shape factor, $\beta$. We then consider several numerical algorithms to simulate the  Langevin equation. The Jeffery orbits can be described using either cartesian or spherical polar coordinates to which we add an appropriate noise term. The analytical solutions of the Fokker-Planck equation for the orientation distribution are validated by comparing with numerical simulations. The steady state solutions depend on the shape factor and the P\'eclet number, defined as the ratio of the diffusion coefficient to the shear rate. We examine the rotation of the particle around the vorticity axis and we calculate the orientational order matrix and the associated scalar order parameters $s$ and $r$ describing nematic ordering and biaxiality, respectively. Using perturbation theory we obtain simple analytical expressions for $s$ and $r$ at large values of the P\'eclet number. For completeness, we also present the solution of the Fokker-Planck equation in two-dimensions.

\section{Jeffery Orbit}

\begin{figure}[t]
 \begin{center}
  \includegraphics[width=9cm]{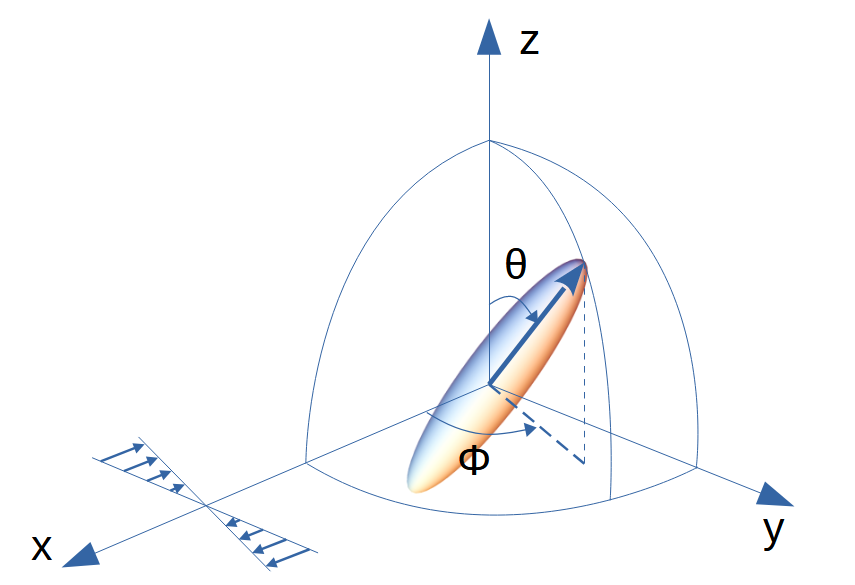}
 \end{center}
   \caption{One of the coordinate systems used to describe the motion of an ellipsoid of revolution. In this case with shear in the $xy$ (v-grad v) plane. The $x,y,z$ axes correspond to the flow, gradient and vorticity directions, respectively. We also used other systems with shear in the $zz$ and $yz$ planes (not shown).}
   \label{fig:coords}
 \end{figure}

The motion of an anisometric particle in a shear flow, without noise, can be described by a Jeffery orbit. A useful starting point is the coordinate-free representation 
\begin{equation}
\dot{\bf p} ={\bf W}{\bf p} +\beta[ {\bf E}{\bf p}-({\bf p}.{\bf E}{\bf p}){\bf p}],
\label{eq:pdot}
\end{equation}
where  ${\bf p}$ is a unit vector specifying the particle orientation, 
${\bf E}=(\nabla {\bf u}+\nabla {\bf u}^T)/2$, ${\bf W}=(\nabla {\bf u}-\nabla {\bf u}^T)/2$ (with the convention $\left(\nabla {\bf u} \right)_{ij}=\partial_{j}u_{i}$) are the strain rate and vorticity tensors and $\beta$ is the shape factor (Bretherton parameter \cite{bretherton1962motion})
\begin{equation}
\beta = \frac{e^2-1}{e^2+1}
\end{equation}
for a particle of elongation $e$. In this article we restrict our attention to the uniaxial, prolate case, $\beta>0$ for which $e$ is the ratio of the longest axis to the shortest one. See Fig. \ref{fig:coords}.
Note that $\beta=1$  and $\beta=0$ correspond to infinitely elongated and spherical  particles respectively.
%, while $\beta=-1$ describes an infinitely thin disk.
Particular realizations of the Jeffery equations have been obtained by choosing the flow geometry. See, for example \cite{jeffery1922motion,leal1971effectA,ishimotoJPSJ2023}. 
For shear in the $xy$ plane, i.e.,  ${\bf u}=\dot{\gamma}y\hat{\bf x}$, substituting in (\ref{eq:pdot}) and using 
standard spherical polar coordinates ($\theta,\;\phi$)

\begin{equation}
\frac{d\theta}{dt} = \frac{{\dot\gamma}\beta}{4}\sin(2\theta)\sin(2\phi)
\end{equation}
and
\begin{eqnarray}
\frac{d\phi}{dt} &=& -{\dot\gamma}\left(\frac{1+\beta}{2}\sin^2\phi+\frac{1-\beta}{2}\cos^2\phi\right)\nonumber\\
&=&-\frac{\dot\gamma}{2}(1-\beta\cos(2\phi)),
\label{eq:phidot3DXY}
\end{eqnarray}
which is independent of $\theta$ (and can describe a stand-alone 2D system: See section \ref{sec:2D}). These equations may be solved exactly:
\begin{equation}\label{eq:phidotexact}
e\tan\phi= - \tan\left(\frac{\dot\gamma t}{e+1/e}+\kappa\right),
\end{equation}
and
\begin{equation}
\tan\theta=\frac{C e}{\sqrt{e^2\sin^2\phi+\cos^2\phi}}.
\end{equation}
where $C$ and $\kappa$ are constants.
The particle rotates in a non-uniform way about the $z$ (vorticity) axis with period 
\begin{equation}
T=\frac{2\pi}{\dot\gamma}\left(e+\frac{1}{e}\right)=\frac{4\pi}{\dot\gamma\sqrt{1-\beta^2}}.
\end{equation}
% Therefore, imposing the shear in the $xy$ plane is the best choice for calculating the mean rate of rotation (see below).
If shear is applied in the $xz$ plane, ${\bf u}=\dot{\gamma}z\hat{\bf x}$,   we get

\begin{eqnarray}
\frac{d\theta}{dt} =&\dot{\gamma}\left(\frac{1}{2}(1-\beta)+\beta\cos^2\theta\right)\cos\phi \label{eq:thetadotxzplane} \\
\frac{d\phi}{dt} =&-\frac{\dot{\gamma}}{2}(1+\beta)\cot\theta\sin\phi \label{eq:phidotxzplane}.
\end{eqnarray}

Note that in this flow the equations for $\dot{\theta}$ and  $\dot{\phi}$ are no longer decoupled. This configuration is, however, the best choice for solving the Fokker-Planck equation (see below).

\section{Langevin Description}

To obtain the Langevin equations we take the equations for the Jeffery orbits given above and add rotational diffusion \cite{Perrin1934,Favro1960,Hubbard1972,McClung1980,mcconnell1980rotational,coffey1996langevin,Coffey2003,junot2019swimming}. As explained in Appendix \ref{appendixStrato}, one gets
\begin{equation}\label{eq:pdotold}
\dot{\bf p} = ({\bf I}-{\bf p}\otimes{\bf p})(\beta {\bf E}+{\bf W}){\bf p}-2D{\bf p}+\sqrt{2D}{\bf p}\times{\bm\xi},
\end{equation}
where ${\bm\xi}$ is a vectorial white noise and the additional term $-2D{\bf p}$ corresponds to the drift induced by the multiplicative noise (transformation from the Stratonovich to the Itô calculus). Note that this drift guarantees the conservation of the norm of the unit vector $\bf p$ when Eq. (\ref{eq:pdotold}) is interpreted in the Itô sense, as we assume in the following.

The Langevin equation can be expressed in Cartesian coordinates. In this case Eq. (\ref{eq:pdotold}) corresponds to randomly displacing the point representing the orientation $p$ in a plane tangent to the surface of the unit sphere and then scaling it back so that it again lies on the surface of the sphere. 

However, in order to compare with the Fokker-Planck equation it is more convenient to express the Langevin equation directly in spherical polar coordinates. 
For pure diffusion the equations are the following \cite{brillinger1997particle,raible2004langevin}:
\begin{eqnarray}
\frac{d\theta}{dt} &=& \frac{D}{\tan\theta}+\sqrt{2D}\xi_1(t)\\
\frac{d\phi}{dt}&=& \frac{\sqrt{2D}}{\sin\theta}\xi_2(t),
\end{eqnarray}
where $\xi_1(t)$ and $\xi_2(t)$ are independent Gaussian white noises.

To perform numerical simulations of these equations we introduce a finite timestep $\delta t$ to calculate the increments:

\begin{eqnarray}
\theta(t+\delta t) &=& \theta(t) + \frac{D}{\tan\theta}\delta t +\sqrt{2D\delta t}\eta_1\\
\phi(t+\delta t)&=& \phi(t)+\frac{\sqrt{2D\delta t}}{\sin\theta}\eta_2,
\end{eqnarray}
where $\eta_1(t)$ and $\eta_2(t)$ are normal variates with mean zero and standard deviation one (${\cal N}(0,1)$).
We can check that this algorithm produces a uniform distribution of particle orientation over
the surface of a sphere. While there are singularities at $\theta=0,\pi$, their presence does not cause serious problems in the numerical simulations.

To simulate the Jeffery orbits with noise, these equations can be combined with those for the Jeffery orbits \cite{junot2019swimming} (the Itô drift being unchanged). For shear in the $xy$ plane we obtain
\begin{widetext}
\begin{eqnarray}\label{eq:thetaphieqShearxy}
\theta(t+\delta t) &=& \theta(t) +\frac{{\dot\gamma}\beta}{4}\sin(2\theta)\sin(2\phi)\delta t + \frac{D}{\tan\theta}\delta t +\sqrt{2D\delta t}\eta_1\\
\phi(t+\delta t)&=& \phi(t)-\frac{\dot\gamma}{2}(1-\beta\cos(2\phi))\delta t +\frac{\sqrt{2D\delta t}}{\sin\theta}\eta_2
\end{eqnarray}
while for shear in the $xz$ plane the result is
\begin{eqnarray}\label{eq:thetaphieqShearxz}
\theta(t+\delta t) &=& \theta(t) +\dot{\gamma}\left(\frac{1}{2}(1-\beta)+\beta\cos^2\theta\right)\cos\phi\;\delta t + \frac{D}{\tan\theta}\delta t +\sqrt{2D\delta t}\eta_1\\
\phi(t+\delta t)&=& \phi(t)-\frac{\dot{\gamma}}{2}(1+\beta)\cot\theta\sin\phi\;\delta t +\frac{\sqrt{2D\delta t}}{\sin\theta}\eta_2
\end{eqnarray}
\end{widetext}

Of course changing the shear plane should not affect the results. For the same parameters $\beta$ and $Pe$ (or $ D / \dot\gamma = 1/Pe)$ we expect to obtain the same results in the two coordinate systems. Examples of Jeffery orbits with noise are given in Fig. \ref{fig:Trajs}.

%fig1

\section{Fokker-Planck description}

Let $\psi(\theta,\phi,t)$ be the probability density function of finding the particle with orientation $(\theta,\phi)$ at time $t$. This satisfies the continuity equation

\begin{equation}
\frac{\partial\psi}{\partial t} = -\boldsymbol{\nabla}\cdot({\bf w}\psi)+\boldsymbol{\nabla}\cdot(D\boldsymbol{\nabla}\psi)
\end{equation}
where $D$ is the rotational diffusion coefficient, assumed constant \cite{burgers1938motion,leal1971effectA}, and where ${\bf w} = ({\bf I}-{\bf p}\otimes{\bf p})(\beta {\bf E}+{\bf W}){\bf p}$, i.e., the deterministic part of the Langevin equation (\ref{eq:pdotold}).

In spherical coordinates, ${\bf w}(\theta,\phi)=(0,\dot\theta,\dot\phi\sin\theta)$ and

\begin{equation}\label{eq:nablawpsi}
\boldsymbol{\nabla}\cdot\left(\bf w\psi\right)=\frac{1}{\sin\theta}\partial_{\theta}\left(\dot{\theta}\sin\left(\theta\right)\psi\right)+\partial_{\phi}\left(\dot{\phi}\psi\right)
\end{equation}

It is particularly convenient to solve the Fokker-Planck equation with shear (v-grad v) in the $xz$ plane, which exploits the higher symmetry of the system with spherical coordinates (the reflection symmetry in the $xz$ plane allows one to use real spherical harmonics). Substituting (\ref{eq:thetadotxzplane}) and (\ref{eq:phidotxzplane}) in (\ref{eq:nablawpsi}), we obtain for the steady state

\begin{equation}
\hat{\Lambda}\psi(\theta,\phi)=\epsilon\nabla^2\psi
\label{eq:FP3D}
\end{equation}
where $\epsilon=D/\dot{\gamma}$ and
\begin{widetext}
\begin{equation}
\hat{\Lambda}=-\frac{3}{2}\beta\sin(2\theta)\cos\phi\;+\left(\frac{1-\beta}{2}+\beta\cos^2\theta\right)\cos\phi\frac{\partial}{\partial\theta}-\frac{1+\beta}{2}\cot\theta\sin\phi\frac{\partial}{\partial\phi}
\end{equation}
\end{widetext}
Doi and Edwards studied a particular case, an infinitely thin needle, $\beta=1$,  for which 
\begin{equation}
\hat{\Lambda}_1=-3\sin\theta\cos\theta\cos\phi\ + \cos^2\theta\cos\phi\frac{\partial}{\partial\theta}-\cot\theta\sin\phi\frac{\partial}{\partial\phi}
\end{equation}
or, in terms of the angular momentum operators:
\begin{eqnarray}
\hat{\Lambda}_1&=\left[\sqrt{\frac{16\pi}{45}}Y_2^0+\frac{1}{3}\right]i\hat{L}_y+\sqrt{\frac{2\pi}{15}}(Y_2^1+Y_2^{-1})\hat{L}_z\nonumber\\
&+3\sqrt{\frac{2\pi}{15}}(Y_2^1-Y_2^{-1})
\end{eqnarray}
where
$\hat{L}_y=i(-\cos\phi\partial/\partial\theta + \cot\theta\sin\phi\partial/\partial\phi)$ and $\hat{L}_z=-i\partial /\partial\phi$.
Here we extend the Doi and Edwards solution to arbitrary values of $\beta$ by noting that the operator can be expressed as
\begin{equation}
\hat{\Lambda}=\beta\hat{\Lambda}_1+\frac{1-\beta}{2}i\hat{L}_y, 
\label{eq:lambdabeta}
\end{equation}
a result that, to the best of our knowledge, has not been given previously.
For $\beta=0$, corresponding to a sphere, Eq. (\ref{eq:FP3D}) simplifies to 
\begin{equation}
i\hat{L}_y\psi(\theta,\phi)=2\epsilon\nabla^2\psi(\theta,\phi)
\end{equation}
whose solution is $\psi(\theta,\phi)=1/4\pi$.
To solve the FP equation for any value of $\beta$, following Doi and Edwards, we introduce the functions
\begin{equation}
S_l^m(\theta,\phi)=\frac{1}{\sqrt{2}}[Y_l^m(\theta,\phi)+(-1)^mY_l^{-m}(\theta,\phi)],
\end{equation}
also known as real spherical harmonics,
which are real and have the property that $S_l^m(\theta,-\phi)=S_l^m(\theta,\phi)$.
The basis set is defined as

\begin{equation}
|lm)=
\begin{cases}
Y_l^m(\theta,\phi),m=0\\
S_l^m(\theta,\phi), m\ne 0
\end{cases}
\end{equation}
with the property
\begin{equation}
(l'm'|lm)=\delta_{ll'}\delta_{mm'}
\end{equation}

We write the distribution function as
\begin{equation}
\psi(\theta,\phi)=\sum_{l\;{\rm even} }^{\infty}\sum_{m=0}^{l}b_{lm}|lm)
\label{eq:psiFP}
\end{equation}
with $b_{00}=1/\sqrt{4\pi}$.

This expansion has the required properties that $\psi(\pi-\theta,\phi+\pi)=\psi(\theta,\phi)$ (even parity) and $\psi(\theta,-\phi)=\psi(\theta,\phi)$ (reflection symmetry in the $xz$ plane).

Substituting Eqs. (\ref{eq:psiFP}) and (\ref{eq:lambdabeta}) in Eq. (\ref{eq:FP3D}) we obtain 

% \frac{D}{\dot{\gamma}}

\begin{equation}
-\epsilon \, l(l+1)b_{lm}=\sum_{l' \ge 0,\;{\rm even}}^ {l_{max}}\sum_{m' =0}^ {l' }(lm|\hat{\Lambda}|l'm' )b_{l'm'}
\label{eq:blm}
\end{equation}
for $l \leq l_{max}$.
From this we generate $l_{max}(1+l_{max}/4)$ simultaneous equations that can be solved to find the coefficients $\{b_{lm}\}$.

For convenience the expression of $(lm|\hat{\Lambda}|l'm' )$ matrix elements are given in Appendix \ref{appendixMatrix}. 
With these results the eigenfunction expansion of the exact distribution for particles with arbitrary shape factor can easily be obtained for any (reasonable) value of $l_{max}$ without the need to evaluate the matrix elements by numerical integration. 

\section{Properties of Interest}

\subsection{Trajectories}

\begin{figure*}[htbp]
 \begin{center}
  \includegraphics[width=17cm]{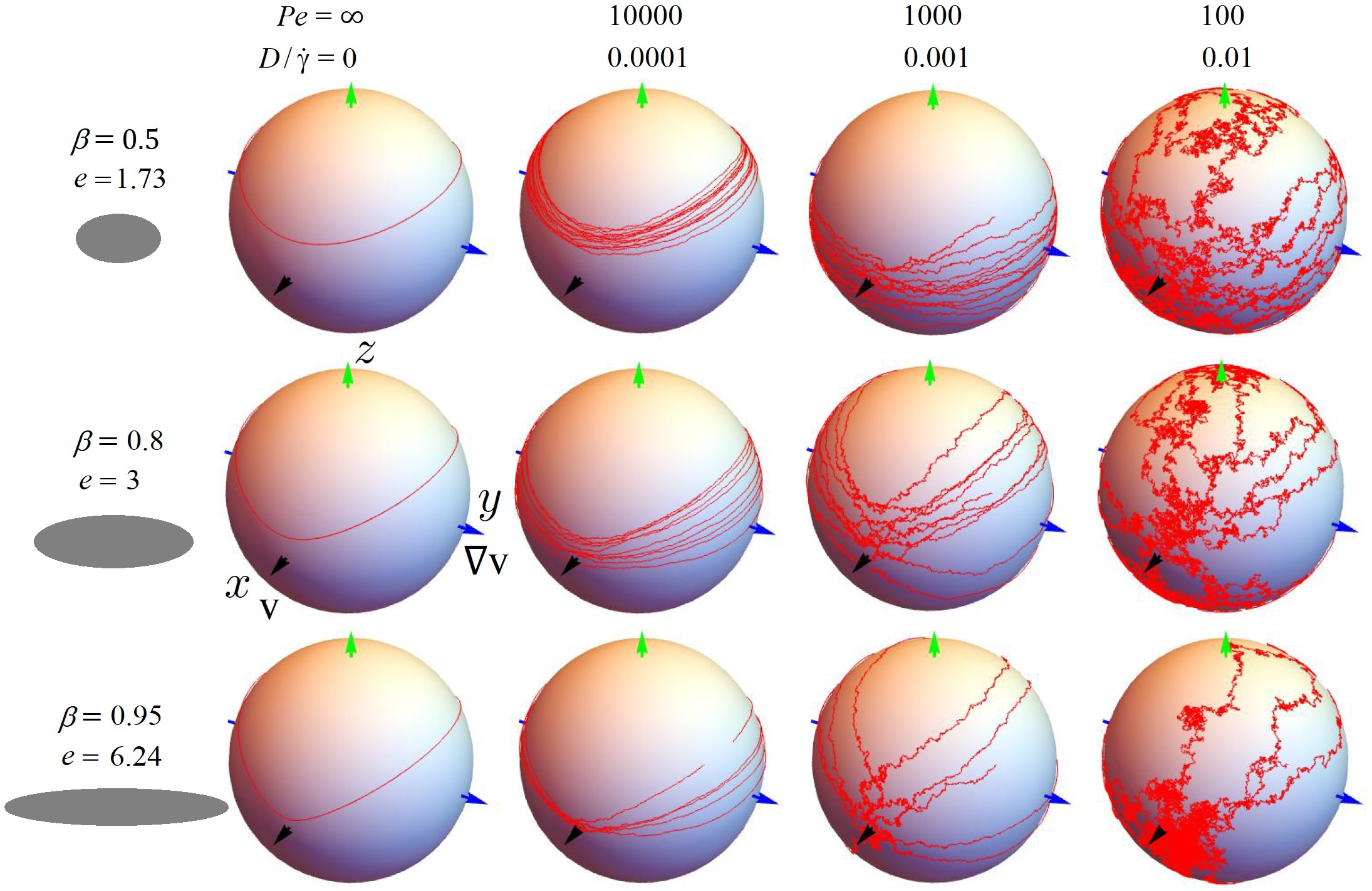}
 \end{center}
   \caption{Trajectory plots with shear v-grad v in the $xy$ plane from numerical simulation for different elongations. In all cases the starting orientation is $\theta=\phi=\pi/3$.}
   %\caption{Trajectory plots with shear (v-grad) v in the $xy$ plane from numerical simulation for $\beta=0.5,0.8,0.95$ left to right panels.    In each panel $D/\dot{\gamma}=0,0.0001,0.001,0.01$ clockwise starting from upper left. In all cases the starting orientation is $\theta=\phi=\pi/3$. The black, blue and green arrows show the $x,y$ and $z$ axes, respectively.}
   \label{fig:Trajs}
 \end{figure*}

Figure \ref{fig:Trajs} illustrates the effect of Brownian motion on Jeffery orbits for particles of shape parameters $\beta=0.5,0.8,0.95$ (corresponding to elongations of $e=1.73,3,6.24$). The 
%fig2
 trajectories were calculated from Langevin simulations using a timestep of 0.001 starting from an orientation of $\phi=\theta=\pi/3$. For zero noise, one recovers the unperturbed Jeffery orbits with their characteristic kayaking motion. For very small noise, $D/\dot{\gamma}=0.0001$, one observes a small drift from the initial orbit, but the general shape remains similar to the unperturbed case. For a ten-fold increase in noise, one starts to see a concentration of the trajectories along the flow direction ($x$) and this effect becomes more pronounced as the particle elongation increases. With a still larger noise of $D/\dot{\gamma}=0.01$ the fractal nature of the trajectories is apparent as they become increasingly aligned with the  flow direction with only occasional excursions to the antipodean orientation.

\subsection{Rotation around the vorticity axis}

We first examine the rotation of the particle around the vorticity axis. For shear in the $xy$ plane, in the steady state we have $\langle d\theta/dt\rangle=0$, where the angular brackets denote a time average, giving 
\begin{equation}
\frac{\dot\gamma\beta}{4}\left\langle\sin(2\theta)\sin(2\phi)\right\rangle+ D\langle\cot\theta\rangle = 0
\end{equation}
and
\begin{equation}
\left\langle\frac{d\phi}{dt}\right\rangle=-\frac{\dot\gamma}{2}(1-\beta\langle\cos(2\phi)\rangle)
\end{equation}
With zero noise, expressing $\phi$ as a function of time using Eq. (\ref{eq:phidotexact}) and integrating over one period we obtain
\begin{equation}\label{eq:fdot3Dzero}
\left\langle\frac{d\phi}{dt}\right\rangle=-\frac{\dot\gamma e}{e^2+1}=-\frac{\dot\gamma}{2}\sqrt{1-\beta^2}
\end{equation}
The rate of rotation increases with decreasing elongation reaching a maximum value of $\dot{\gamma}/2$ for spherical particles, $\beta=0$. Infinitely thin particles, $\beta=1$ are singular in that they do not rotate periodically. Indeed $\phi=0$ is an attractor of the differential equation (\ref{eq:phidot3DXY}): the particle eventually aligns with the flow for all starting orientations.

%%Interestingly this implies that the infinitely thin particles, $\beta=1$, do not rotate on average, while a spherical 
%%particle, $\beta=0$, rotates at $\dot{\gamma}/2$. 

\begin{figure}
 \begin{center}
  \includegraphics[width=8.5cm]{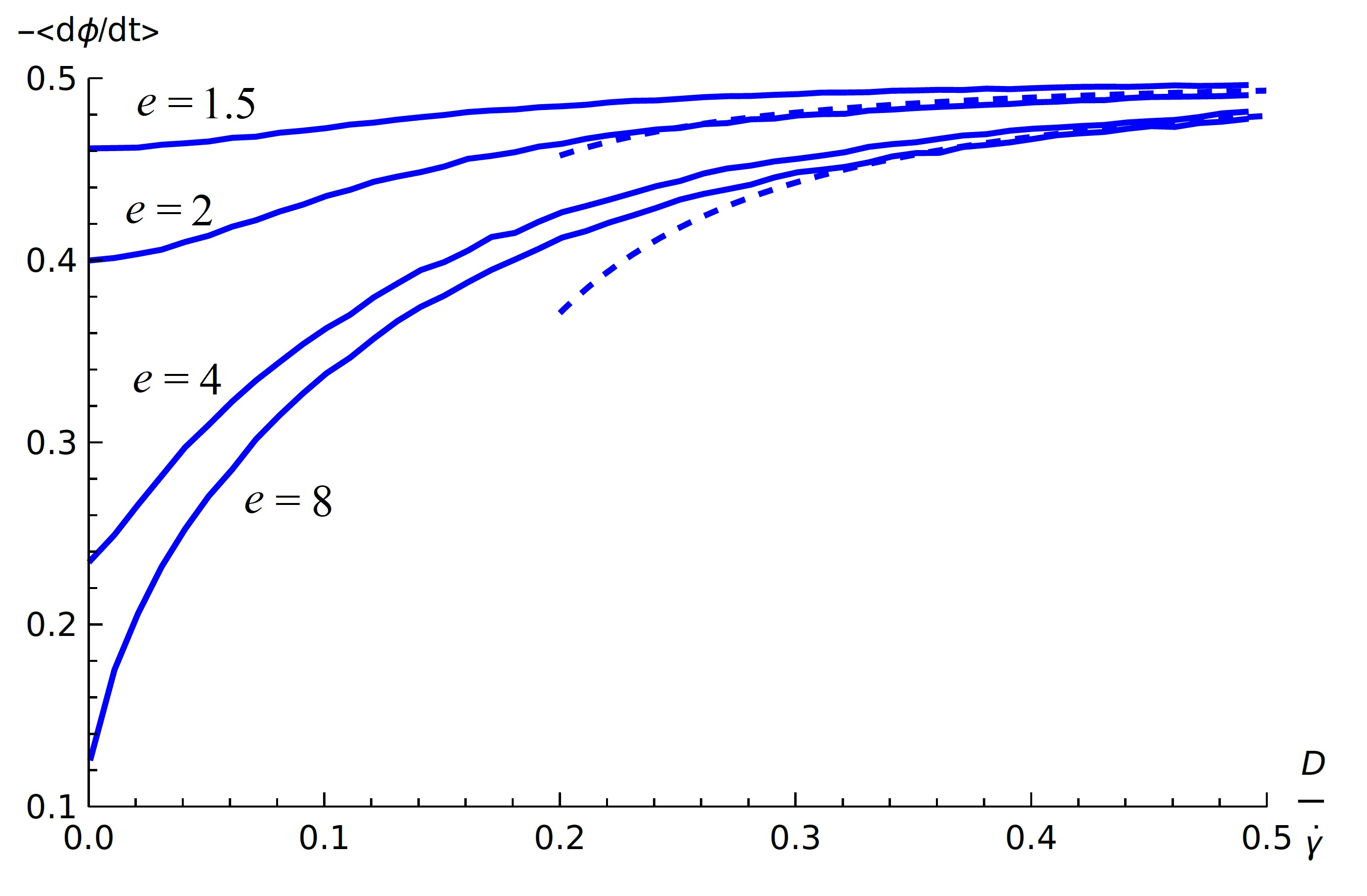}
 \end{center}
   \caption{Mean angular velocity for particle elongations $e=1.5,2,4,8$, 
   %top-to-bottom left hand side, 
   calculated from simulations with shear in the $xy$ plane. The limiting values for small $D$ are 
   correctly given by Eq. (\ref{eq:fdot3Dzero}). The dashed lines show the results of perturbation theory to second order (Appendix \ref{appendixPertPe}). }
   \label{fig:omegabar3D}
 \end{figure}

The effect of noise on the rate of rotation is shown in Fig. \ref{fig:omegabar3D}. As in the noiseless case, for a fixed level of noise the rate of rotation increases with decreasing elongation and, for fixed elongation ($\beta$), the rate of rotation increases with increasing noise. Perturbation theory, as developed in Appendix \ref{appendixPertPe}, can be used to calculate the rotation speed in the limit of large noise (or small Pe). 

%fig3

\subsection{Orientational distribution function} 

\begin{figure}[h!]
 \begin{center}
  \includegraphics[width=\columnwidth]{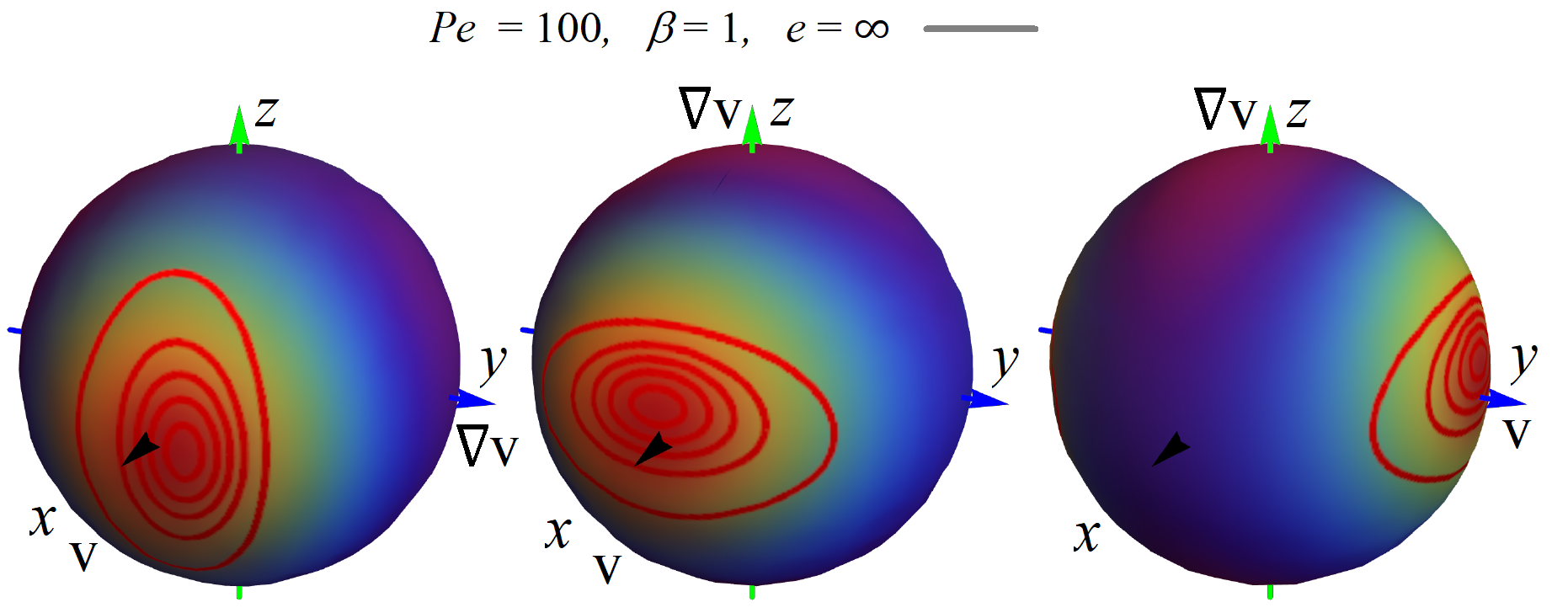}
 \end{center}
   \caption{Density plots of the particle orientation from Langevin simulations with v-grad v in the $xy$, $xz$ and $yz$ planes for $Pe=100$ and $\beta=1$. The contours show the solution of the Fokker-Planck equation (\ref{eq:psiFP}) with $l_{max}=16$. %The black, blue and green arrows show the $x,y$ and $z$ axes, respectively.
   }
   \label{fig:JOXYXZYZ}
 \end{figure}

\begin{figure*}[t]
 \begin{center}
  \includegraphics[width=\textwidth]{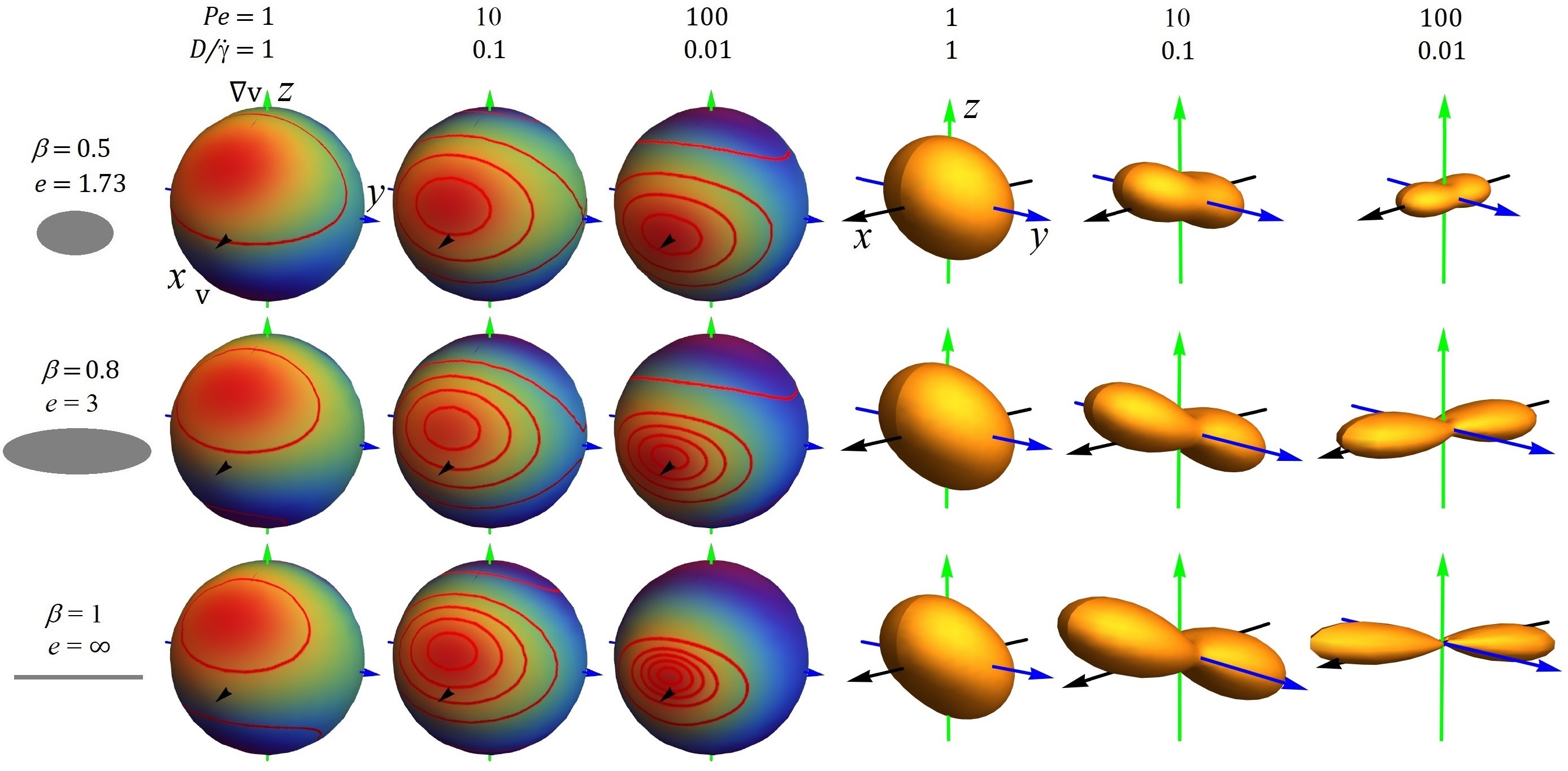}
 \end{center}
   \caption{Left: Density plots of the particle orientation calculated from Langevin simulations with v-grad v in the $xz$ plane. The contours show the solution of the Fokker-Planck equation (\ref{eq:psiFP}) with $l_{max}=16$, right: The same Fokker-Planck solutions shown as probability surfaces. The distance of the surface from the origin in a  given direction is proportional to the probability of the particle orientation in that direction. %\textbf{Note that the viewing angle can be misleading in some cases. For instance, the probability surface for the case $\beta = 1$ and $Pe = 1$ might give the impression that the maximum lies along the $y$ axis whereas it is actually located around $45$ degrees in the $xz$ plane.}
   }
   \label{fig:Probsurface9a}
 \end{figure*}

We present probability density functions of  the particle orientation obtained from Langevin simulations as heat maps in Figs. \ref{fig:JOXYXZYZ}  and \ref{fig:Probsurface9a}. Figure \ref{fig:JOXYXZYZ} is for a needle, $\beta=1$, with $Pe=100$ for three perpendicular v-grad v planes, $xy$, $xz$ and $yz$. 
%fig4
At this relatively high P\'eclet number  the orientational distribution is strongly anisotropic. For the case with shear in the $xy$ plane the peak of the distribution is centered in the same plane.  It does not point directly along the $x$ axis, but is rotated towards the $y$-axis \cite{marschallPRE2020}. The spot is not symmetric, being somewhat stretched in the $z$ direction. It is clear that the remaining distributions $xz$ and $yz$ are related to the first one by a simple rotation. This helps to confirm the correct functioning of the Langevin algorithm as well as  the graphical representation.

In Fig. \ref{fig:Probsurface9a} (left) we show the effect of varying the P\'eclet number between 1 and 100 for particles with shape parameters $\beta=0.5$, $0.8$ and $1$. As expected, with increasing noise (decreasing $Pe$) the distributions spread. For a fixed value of $Pe$, the distribution becomes more diffuse as the particle becomes less elongated. We also show the solutions of the Fokker-Planck equation (\ref{eq:psiFP}) as contours superimposed on the density plots. Fig. \ref{fig:Probsurface9a} (right) shows the same solutions as probability surfaces.  The solution was obtained with shear in the $xz$ plane.   The full symmetry of the distributions is more clearly apparent in this probability surface representation. 

%fig5

\subsection{Order Matrix}
The orientational order can be characterized with the order matrix \cite{majumdar2010equilibrium}
\begin{equation} \label{eq:QMpsi}
\mathbf{Q}=\int_{S^2} \left(\mathbf{p}\otimes \mathbf{p}- \frac{\mathbf{I}}{3}\right)\psi(\mathbf{p})d\mathbf{p}
\end{equation}
where $\mathbf{I}$ is the unit matrix and $\mathbf{p}=(\sin\theta\cos\phi,\sin\theta\sin\phi,\cos\theta)$. It follows 
from the definition that $\mathbf{Q}$ is a symmetric, traceless $3\times 3$ matrix. It can be expressed in terms of a triad of 
orthonormal vectors
\begin{equation}
\mathbf{Q}=\lambda_1 \mathbf{e_1}\otimes\mathbf{e_1} + \lambda_2 \mathbf{e_2}\otimes\mathbf{e_2} +\lambda_3 \mathbf{e_3}\otimes\mathbf{e_3} 
\end{equation}
with $\lambda_1+\lambda_2+\lambda_3=0$.  Three kinds of order are possible: (i) isotropic with all eigenvalues equal to zero; (ii) uniaxial with one positive eigenvalue and two equal negative eigenvalues (or two equal positive and one negative, where the preferred orientation is along a plane instead of along a direction. We observe the latter near flat walls for rod-like particles) and (iii) biaxial with three distinct eigenvalues. The eigenvalues satisfy $-1/3\le\lambda_i\le2/3$. 

The Q-matrix may be represented as
\begin{equation} 
\mathbf{Q}=s\left(\mathbf{e_1}\otimes \mathbf{e_1}- \frac{\mathbf{I}}{3}\right)+r\left(\mathbf{e_2}\otimes \mathbf{e_2}- \frac{\mathbf{I}}{3}\right)
\end{equation}
with the scalar order parameters $(s,r)$ describing nematic ordering  and biaxiality, given by 
\begin{equation}
s = 2\lambda_1+\lambda_2,\;\;
r = 2\lambda_2+\lambda_1,
\end{equation}
assuming that the eigenvalues are labeled so that $\lambda_1\ge\lambda_2\ge\lambda_3$. (An alternative representation of the $Q$ tensor \cite{majumdar2010landau} gives rise to another set of parameters, $S=3\lambda_1/2$, the widely used uniaxial nematic order parameter, and $T=(2\lambda_2+\lambda_1)/2$, an alternative measure of biaxiality. The two sets are related by $S=s-r/2,T=r/2$. For  a uniaxial system there is no difference, $s=S$).

\begin{figure}[h]
 \begin{center}
  \includegraphics[width=8.5cm]{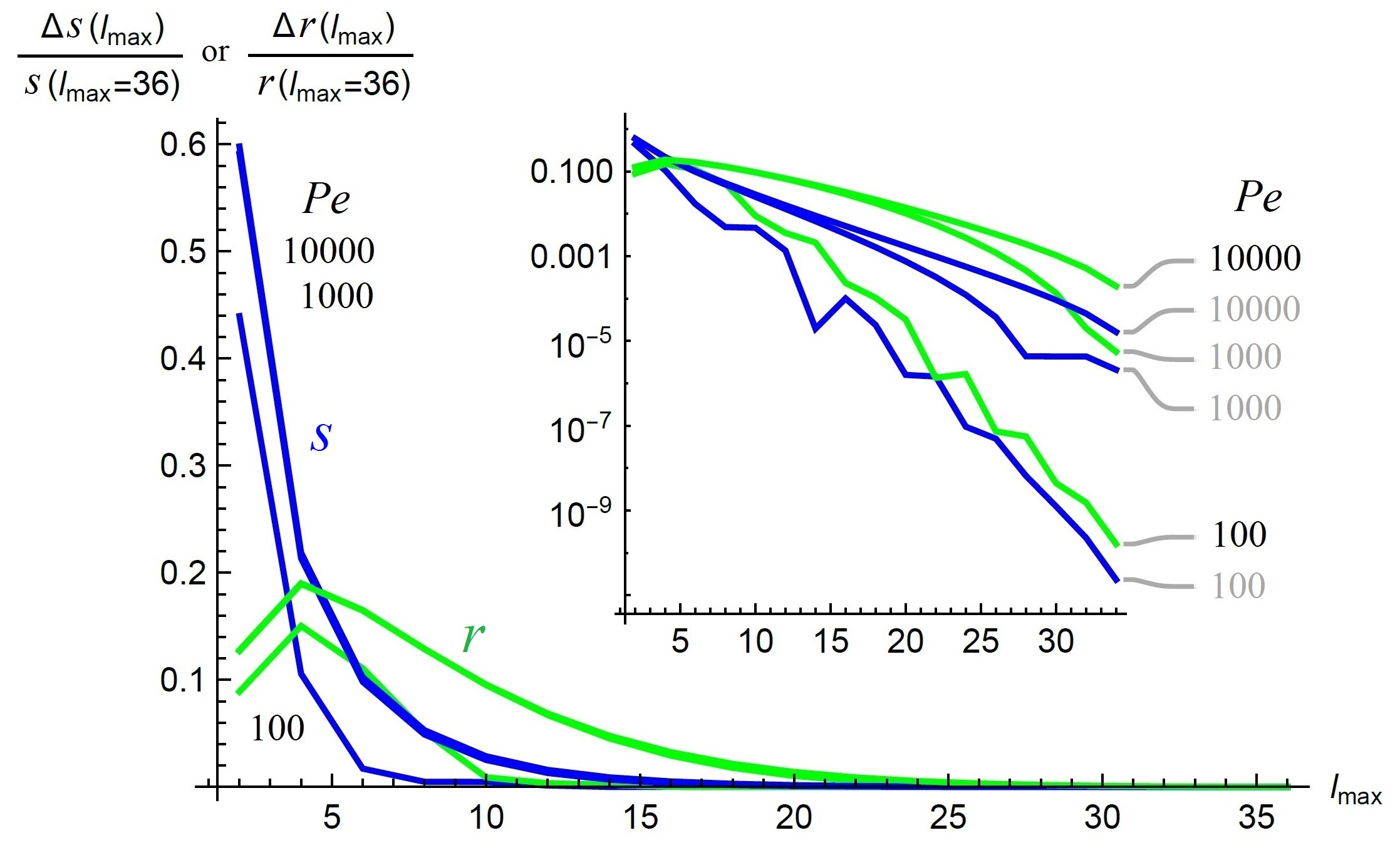}
 \end{center}
   \caption{Convergence of the scalar order parameters $s$ (blue) and $r$ (green) as a function of $l_{\rm max}$ for $Pe=100,1000,10000$, curves from  bottom to top, respectively.}
   \label{fig:convergenceOPS}
 \end{figure}

In the simulations the order tensor is evaluated by computing the tensor product $\mathbf{p}\otimes \mathbf{p}$ averaged over the  time steps. We can calculate it from the Fokker-Planck equation by substituting the orientational probability function, Eq. (\ref{eq:psiFP}), in Eq. (\ref{eq:QMpsi}) to obtain 
%\begin{widetext}
\begin{equation}
\mathbf{Q}=
\begin{pmatrix}
\sqrt{\frac{4\pi}{45}}(\sqrt{3}b_{22}-b_{20}) & 0&-\sqrt{\frac{4\pi}{15}}b_{21}  \\
0 &  -\sqrt{\frac{4\pi}{45}}(\sqrt{3}b_{22}+b_{20})& 0 \\
-\sqrt{\frac{4\pi}{15}}b_{21} & 0 & 2\sqrt{\frac{4\pi}{45}}b_{20}
\end{pmatrix},
\end{equation}
%\end{widetext}
indicating that the order matrix  depends only on the second moments of the distribution $b_{20},b_{21},b_{22}$. To obtain these one chooses a value of $l_{\rm max}$ and then constructs the $l_{\rm max}(1+l_{\rm max}/4)/2$ simultaneous equations (\ref{eq:blm}) that are solved for $b_{lm}$. The values of $b_{20},b_{21},b_{22}$ depend on $Pe$ and $l_{\rm max}$: see Fig. \ref{fig:convergenceOPS}. We typically used $l_{\rm max}=24-36$ in the results presented here.

We also use perturbation theory to obtain analytical expressions for $s$ and $r$ valid at 
low noise. This involves expanding the orientational distribution function as a power series in either $\beta$ or $Pe$. With the former we obtain $s=2r=\frac{2\beta}{5\sqrt{1+36(D/\dot{\gamma})^2}}$ while the latter yields $s=2r=\beta Pe/15$. Details of the calculations are presented in Appendices C and D.

%fig6

%fig7
\begin{figure*}
\begin{center}
   \includegraphics[width=14cm]{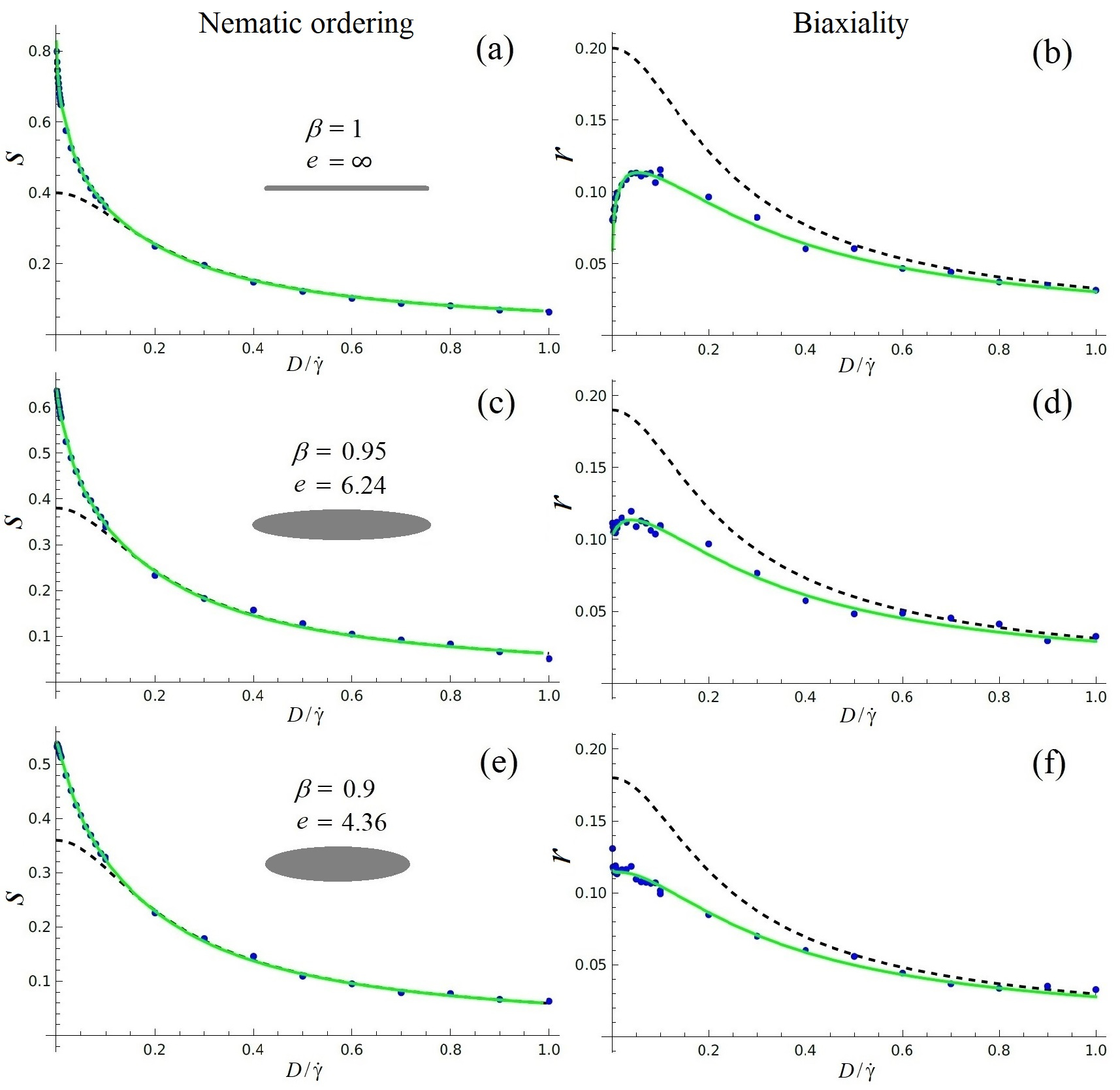}
\end{center}
%\subfloat[$\beta=1$]{\includegraphics[width = 3in]{Figures/NOPbetaOneAnglesJoinT001.pdf}} 
%\subfloat[$\beta=1$]{\includegraphics[width = 3in]{Figures/BOPbetaOneAnglesJoinT001.pdf}}\\
%\subfloat[$\beta=0.95$]{\includegraphics[width = 3in]%{Figures/NOPbeta95AnglesJoinT0001Longer.pdf}}
%\subfloat[$\beta=0.95$]{\includegraphics[width = 3in]{Figures/BOPbeta95AnglesJoinT0001Longer.pdf}} \\
%\subfloat[$\beta=0.9$]{\includegraphics[width = 3in]{Figures/NOPbeta90AnglesJoinT0001Longer.pdf}}
%\subfloat[$\beta=0.9$]{\includegraphics[width = 3in]{Figures/BOPbeta90AnglesJoinT0001Longer.pdf}} 
\caption{Scalar order parameters $s$ and $r$ as a function of $D/\dot{\gamma}$. FP equation (green); $\beta$ perturbation theory (dashed); Langevin simulations $xz$ (blue).}
\label{fig:OParams}
\end{figure*}

Figure \ref{fig:OParams} shows the scalar order parameters $s$ and $r$ for $\beta=1,0.95,0.9$ calculated from the simulations, the series solution of the FP equation (using $l_{max}=24$)  and $\beta$-perturbation theory for  $0.01\le D/\dot{\gamma}\le 1$. The simulations were performed for $5\times 10^6$ steps with a timestep of $\delta t=0.001$. We verified that the simulation results do not depend on the choice of the shear plane. For both $s$ and $r$ the simulation results are in agreement with the FP equation for a wide range of $D/\dot{\gamma}$. 
The value of $r$ (biaxiality) computed from the simulations displays larger fluctuations than the value of $s$. Interestingly, both the solution of the FP equation and simulation exhibit a maximum value of $r$ for $D/\dot{\gamma}\approx 0.05$ for $\beta>0.9$. Perturbation theory provides an excellent description of the nematic ordering $s$ for $D/\dot{\gamma}> 0.1$. It does not work as well as for describing the biaxiality ($r$ does not exhibit a maximum), but does approach the simulation results as $D/\dot{\gamma}$ increases.

The overall decreasing trend of $s$ with $D/{\dot \gamma}$ is what we would expect: noise generally reduces nematic ordering, leading to smaller values of $s$. 
Regarding the biaxiality the situation is more complex: for large noise $r$ is expected to 
decrease with noise as in the large noise limit the system is expected to become isotropic.
In the limit of small noise however, where the value of $s$ is increasing and for very elongated particles it converges to 1 the value of $r$ should converge to 0 by definition. 
This is what we see in the small noise limit in Figs. \ref{fig:OParams}(b,d).

\begin{figure}[h]
 \includegraphics[width=8.5cm]{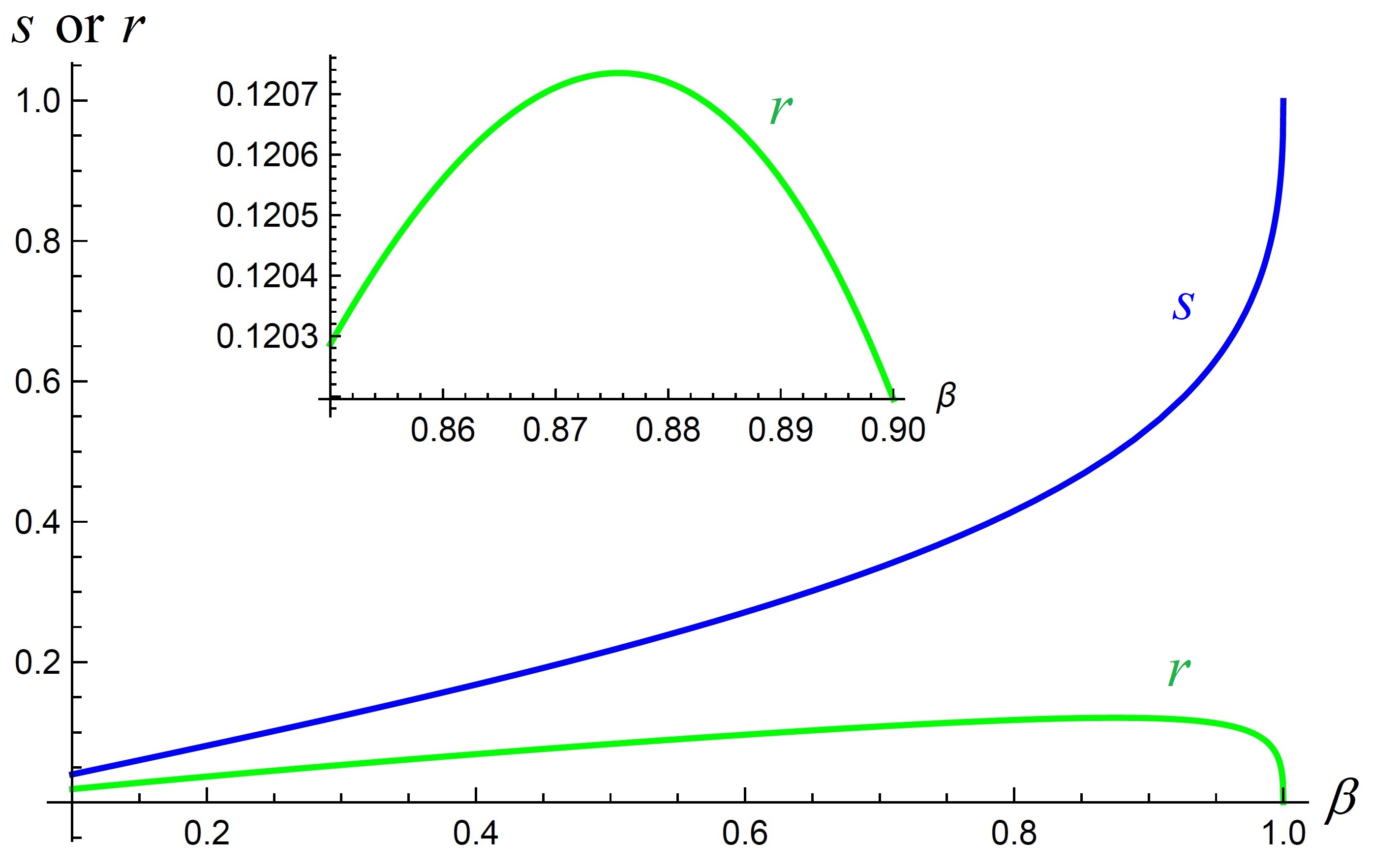}
\caption{Scalar order parameters  $s$ (blue) and $r$ (green) of Jeffery orbits without noise averaged over the initial (isotropic) orientation as a function of $\beta$.}
\label{fig:OPSZero}
\end{figure}

We analyzed the behavior of the nematic ordering and the biaxiality for Jeffery orbits without noise using the procedure outlined in Appendix \ref{appendixOPNoNoise}. The results are shown in Fig. \ref{fig:OPSZero}. While the value of $s$ increases more and more rapidly as $\beta$ approaches one, $r$ displays a maximum at $\beta=0.876$ before rapidly decreasing to zero.

%fig8

\section{In- and out-of plane distributions}

In addition to the full orientational distributions described above, it is useful to examine the angular distribution of the particle in the v-grad v plane as well as in the perpendicular (vorticity) direction. 
This can be done most easily using the simulation algorithm with the shear in the $(x,y)$ plane. In this case the orientation of the particle in the v-grad v plane is specified by $\phi$ and the orientation in the perpendicular plane is given by $\delta=\pi/2-\theta$. The corresponding distributions, denoted by $f_{\parallel}(\phi)$ and $f_{\perp}(\delta)$  respectively, are the marginal distributions of the full orientational distribution: 

\begin{equation}
f_{\parallel}(\phi) = \int_0^\pi \; \psi(\theta,\phi) \sin\theta d\theta;\;\; f_{\perp}(\delta) = \int_0^{2\pi} \; \psi(\delta,\phi) d\phi
\end{equation}

\begin{figure*}[t]
 \begin{center}
  \includegraphics[width=17cm]{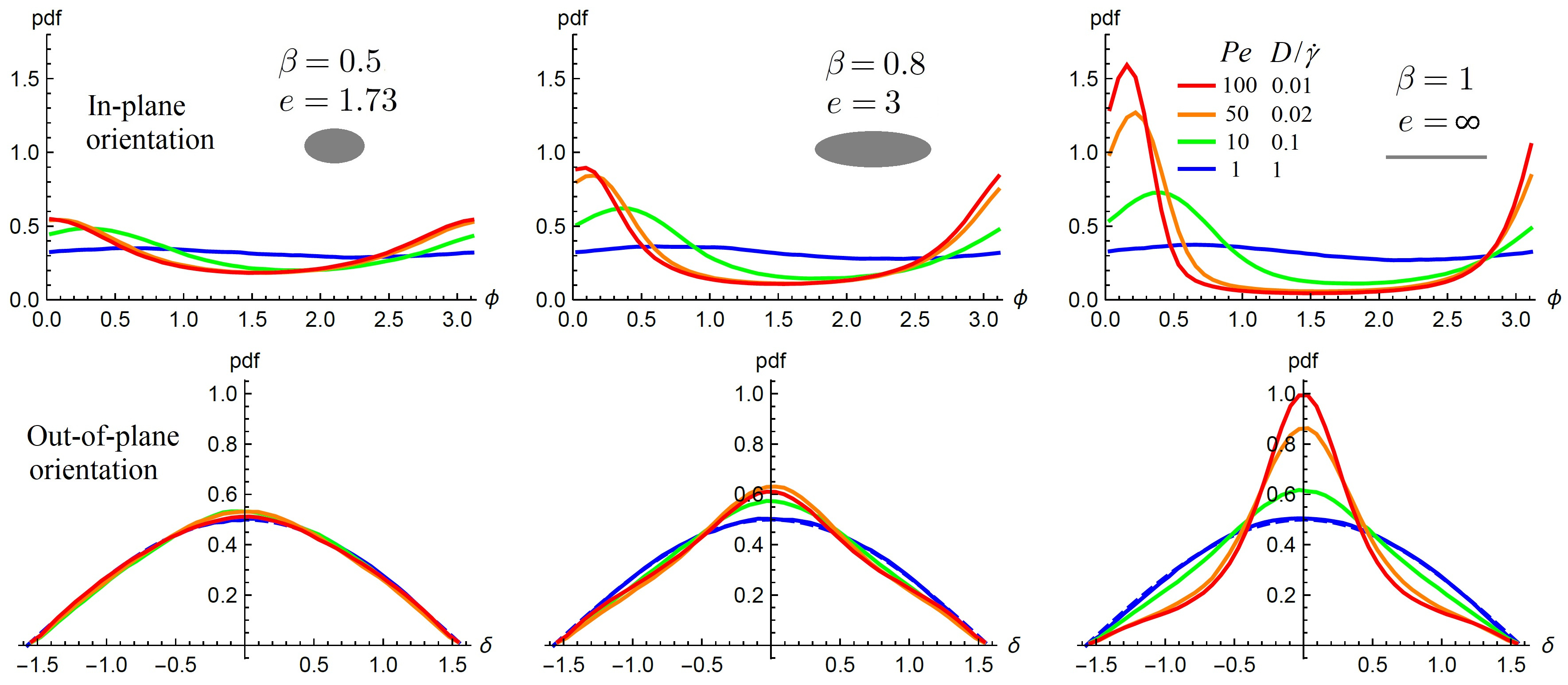}
 \end{center}
   \caption{In-plane (upper row) and out-of-plane (bottom row) orientational distributions for $Pe=1,10,50,100$, for three different particle elongations, calculated from Langevin simulations with shear in the $xy$ plane. 
   %The most peaked distributions correspond to the highest P\'eclet number and the left, middle and right columns correspond to $\beta=0.5,0.8,1$. 
   The dashed curve shows the out-of-plane distribution ($\cos\delta$) for a system with random orientation}.
   \label{fig:InOut}
 \end{figure*}

Results for three different Bretherton parameters, $\beta=0.5,0.8,1$ (corresponding to elongations $e=\sqrt{3},3,\infty$), each for $Pe=1,10,50,100$, are shown in Fig. \ref{fig:InOut}.

%fig9

For large noise ($Pe=1$), the out of plane distribution $f_{\perp}(\delta)$ is basically equal to the uniform distribution ($\cos\delta$) for all three particle shapes. 
The in-plane distribution $f_{\parallel}(\phi)$ is not perfectly uniform. Rather, it has a slight modulation, which has similar amplitude for the three types of grains. The peak of the distribution is located at around 45 degrees, which is in accordance with 2D calculations by Marschall et al. \cite{marschallPRE2020} and other data \cite{lettinga2004non}.

For the least elongated particle ($\beta=0.5$) decreasing the amplitude of the noise results in only a minor change in $f_{\perp}(\delta)$, while  $f_{\parallel}(\phi)$ changes significantly.
Decreasing noise leads to a sharper maximum and a shift of the maximum towards 0.
The out-of-plane distribution is very similar to a uniform distribution, irrespective of the amplitude of the noise.
This means that decreasing noise does not drive the particles into the v-grad v plane.

For a longer particle ($e=3$) both distributions $f_{\perp}(\delta)$  and $f_{\parallel}(\phi)$ change with decreasing amplitude of the noise. 
The in-plane distribution $f_{\parallel}(\phi)$ becomes notably sharper than for $e=1.73$. The average angle appears to be larger for $e=3$ than for $e=1.73$, as observed by Marschall et al. \cite{marschallPRE2020}.
The peak of $f_{\perp}(\delta)$ also becomes larger, meaning that for smaller noise the $e=3$ particles have a greater tendency to remain in the proximity of the v-grad v plane than the $e=1.73$ particles. 

For very long particles ($e=\infty$; $\beta=1$) decreasing noise results in sharper peaks for both the in-plane and out-of-plane orientation distributions. This means that decreasing noise not only leads to strong ordering in the v-grad v plane, but the particles are also clearly driven towards the the v-grad v plane.
%Finally we note that there is little difference between the results for $e=5$ and $e=\infty\;(\beta=1)$ particles.

\begin{figure}[htbp]
 \begin{center}
  \includegraphics[width=8.5cm]{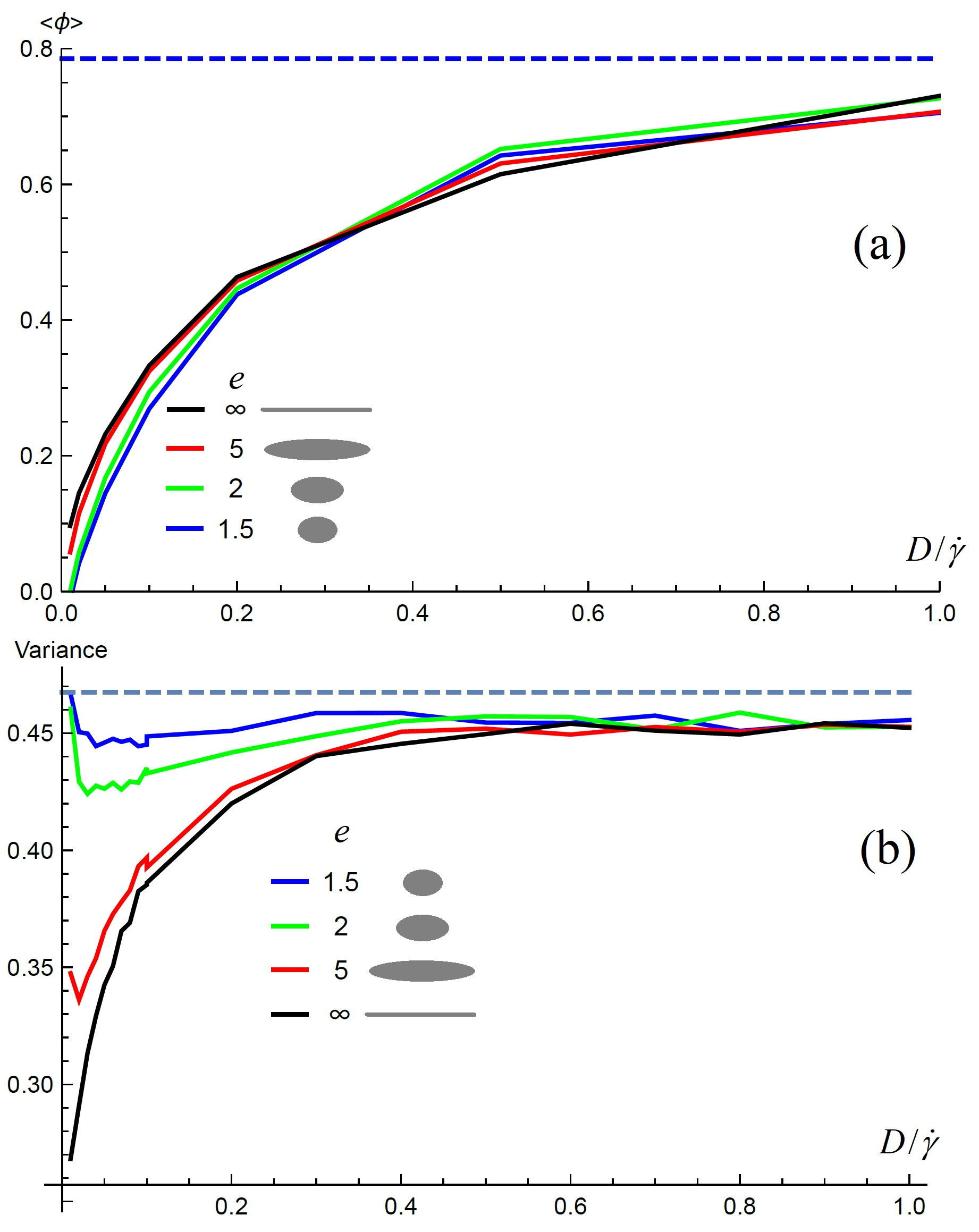}
 %\end{center}
%\includegraphics[width = \columnwidth]{Figures/CircularMeanGrapha.pdf} 
%\includegraphics[width = \columnwidth]{Figures/CircularVarianceGraphA.pdf}
\end{center}
% \subfloat[$\beta=0.5$]{\includegraphics[width = 3in]{Figures/NOPbeta05.pdf}}
% \subfloat[$\beta=0.5$]{\includegraphics[width = 3in]{Figures/BOPbeta05.pdf}} 
\caption{(a) Circular mean of the in-plane orientation for $e=1.5,2,5,\infty\; (\beta=1)$. The dashed line is $\pi/4$; (b) Variance of out-of-plane orientation for $e=1.5,2,5,\infty\; (\beta=1)$.  The dashed line is the value for a random distribution, $(\pi^2-8)/4$. }
\label{fig:MeanVariance2D}
\end{figure}

The circular mean of the in-plane orientation, $\langle\phi\rangle_{CM}$, is shown as a function of $D/{\dot \gamma}=1/Pe$ in Fig. \ref{fig:MeanVariance2D}(a). The behavior is similar to the result obtained by Marschall et al. \cite{marschallPRE2020} for a strictly two-dimensional system, that is the mean angle increases with noise ($D/{\dot \gamma}$) and tends to an asymptotic value of $\pi/4$. For a given value of $D/{\dot \gamma}$ the mean angle increases with increasing particle elongation. 
%Note that for values of $D/{\dot \gamma}$ closer to one, the fluctuations (not yet shown) are large. 

For the out-of-plane distribution the mean angle is always zero (no symmetry breaking). We show the (standard) variance, $\langle\delta^2\rangle$ as a function of   $D/{\dot \gamma}=1/Pe$ in Fig. \ref{fig:MeanVariance2D}(b). All the curves tend to the value corresponding to a uniform distribution $(\pi^2-8)/4$ as the noise increases. For the particle of elongation $e=1.5$ the distribution is nearly uniform for all values of the noise and, consequently, the variance shows little variation. For $e=2$ the variance is a non-monotonic function of $D/{\dot \gamma}$, with a minimum value at about $D/{\dot \gamma}=0.5$.  More elongated particles display a strongly peaked distribution, yielding a small variance, for small values of $D/{\dot \gamma}$.

% Unfortunately, it is not easy to calculate these distributions using the Fokker-Planck equation, at least in the way that we have set it up with $xz$ as the v-grad v plane and $y$ as the vorticity axis. Let $\gamma$ and $\delta$ denote the in-plane and out-of-plane angles. Then one can show that $\cos\delta=\sin\theta\sin\phi$ and $\cos\gamma=\sin\theta\cos\phi/\sqrt{\sin^2\theta \cos^2\phi+\cos^2\theta}$. It seems that we cannot get the distribution of $\gamma$ and $\delta$ analytically. The best we could do would be to sample $\theta,\phi$ from $\psi(\theta,\phi)$, which itself is not an easy task, and then calculate the corresponding values of $\gamma$ and $\delta$. But this offers no advantage compared to direct Langevin simulation.

%fig10

\section{Two dimensions}\label{sec:2D}

For completeness we now present results for the purely two-dimensional motion of an anisometric particle in a shear flow. This case has recently been studied by Marschall et al. \cite{marschallPRE2020} using numerical solutions of the Langevin equation. Here we present the solution of the corresponding Fokker-Planck equation.

The overdamped Langevin equation for the orientation of the particle at time $t$  can be written as
\begin{equation}
\frac{d\phi}{dt} = f(\phi) + \sqrt{2D}\xi,
\label{eq:FP2D}
\end{equation}
where $\xi$ is Gaussian white noise and
\begin{equation}
f(\phi)=-\frac{1}{2}\dot{\gamma}(1-\beta\cos2\phi),
\end{equation}
where $\beta$ has the same definition as in 3D. Moreover, due to the decoupling of $\phi$ and $\theta$ in 3D for v-grad v in the $xy$ plane, the form of $f(\phi)$ is the same as in Eq.~(\ref{eq:phidot3DXY}).
% with $b=(e^2-1)/(e^2+1)$ and $\dot{\gamma}>0$ is the shear rate. 
Note that with this choice, the major axis of the ellipse is parallel to the flow when $\phi=0$. The sign of the function $f(\phi)$ is chosen so that the particle rotates in the clockwise sense. 

The corresponding Fokker-Planck equation for the orientational distribution, $\psi(\theta,t)$, is
\begin{equation}\label{eq:FP}
\frac{\partial \psi(\theta,t)}{\partial t}=-\frac{\partial}{\partial\phi}[f(\phi)\psi(\phi,t)]+\epsilon\frac{\partial^2\psi(\phi,t)}{\partial\phi^2}.
\end{equation}

Let us first focus on the steady state for which
\begin{equation}\label{eq:TE}
\frac{d}{d\phi}[f(\phi)\psi(\phi)]=\epsilon\frac{d^2\psi}{d\phi^2}.
\end{equation}
Integrating once we obtain
\begin{equation}\label{eq:TE1}
f(\phi)\psi(\phi)=\epsilon\frac{d\psi}{d\phi}+c
\end{equation}
If there is no noise $D/\dot{\gamma}=0$ and
\begin{equation}
f(\phi)\psi(\phi) = c.
\end{equation}
This simple equation expresses the fact that the probability to find the particle at a given angle is inversely proportional to its instantaneous angular velocity. 

From the normalization condition 
\begin{equation}
\int_{-\pi/2}^{\pi/2}d\phi\;\psi(\phi) = 1
\end{equation}
we find that $c=\dot{\gamma}\frac{e}{\pi(e^2+1)}$, so
\begin{equation}\label{eq:psi0}
\psi_0(\phi)=\frac{e}{\pi(1+(e^2-1)\sin^2\phi)}.
\end{equation}
This is a symmetric function, $\psi_0(-\phi)=\psi_0(\phi)$ with a maximum value of $e/\pi$ at $\phi=0$ and a minimum value of $1/(e\pi)$ at $\phi=\pm\pi/2$. It is also independent of the shear rate. 

Various approaches can be used to solve Eq. (\ref{eq:FP}) including expansion in circular harmonics, a direct method and singular perturbation theory. We focus on the first as it corresponds to method used in 3D. The direct solution only works in 2D where it gives the same results as the circular expansion method. Perturbation theory is more limited, but it can give information in the limit of small noise.

\subsection{Solution using circular harmonic expansion}

Since $\psi(\phi)$ is a periodic function on the interval $-\pi/2<\phi<\pi/2$ it may be expanded in circular harmonics:

\begin{equation}\label{eq:harexp}
\psi(\phi)=\sum_{k=0}a_k\cos(2k\phi)+b_k\sin(2k\phi),
\end{equation}
that clearly satisfies $\psi(-\pi/2)=\psi(\pi/2)$.
The normalization condition requires that $a_0=1/\pi$. Also $b_0=0$. Given the distribution function the coefficients can be obtained as
\begin{eqnarray}
a_k&=&\frac{2}{\pi}\int_{-\pi/2}^{\pi/2}\cos (2k\theta) \psi(\theta)d\theta\nonumber\\
b_k&=&\frac{2}{\pi}\int_{-\pi/2}^{\pi/2}\sin (2k\theta) \psi(\theta)d\theta.
\end{eqnarray}

Let us first apply this to the known solution in the absence of noise, $\psi_0(\theta)$.  Since this function is symmetric $b_k=0$ for all $k$. The coefficients of the cosine terms are obtained from
\begin{equation}
a_m=\frac{2}{\pi}\left(\frac{e-1}{e+1}\right)^m, m\ge 1.
\end{equation}
Now to obtain the solution in the presence of noise we write the FP equation in the form
\begin{equation}\label{eq:TE1}
\frac{d}{d\phi}[(1-\beta\cos 2\phi)\psi(\phi))]=-2\epsilon\frac{d^2\psi}{d\phi^2},
\end{equation}
where we have used the relation $\sin^2\phi=\frac{1}{2}(1-\cos2\phi)$. Substituting Eq. (\ref{eq:harexp}) in Eq. (\ref{eq:TE1}),
expressing the result entirely in terms of the basis functions $\cos2k\phi$ and $\sin2k\phi$ and equating their coefficients we find
\begin{equation}
2b_k-\beta(b_{k-1}+b_{k+1})=8\epsilon ka_k
\end{equation}
and
\begin{equation}
-2a_k+\beta((1+\delta_{k,1})a_{k-1}+a_{k+1})=8\epsilon kb_k
\end{equation}
for the cosine and sine terms respectively, where $\delta_{k,1}$ is the Kronecker delta.
%
%fig11
\begin{figure}[htbp]
 \begin{center}
  \includegraphics[width=8.5cm]{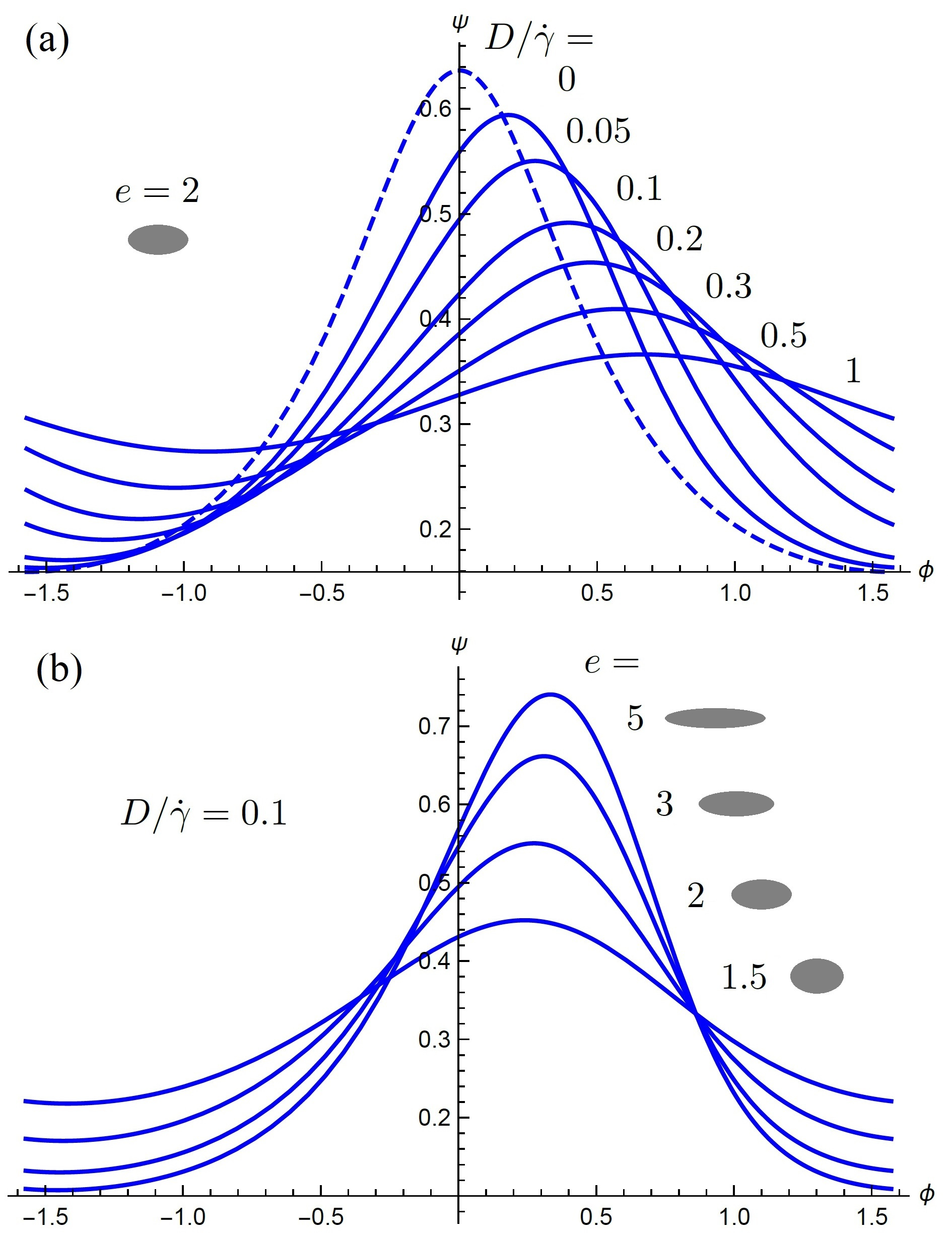}
 \end{center}
%\includegraphics[width = \columnwidth]{Figures/distributions2D.pdf} 
%\includegraphics[width = \columnwidth]{Figures/distributions2DC.pdf}
%\end{center}
% \subfloat[$\beta=0.5$]{\includegraphics[width = 3in]{Figures/NOPbeta05.pdf}}
% \subfloat[$\beta=0.5$]{\includegraphics[width = 3in]{Figures/BOPbeta05.pdf}} 
\caption{Orientational distributions in 2D. (a) $e=2$ and $D/\dot{\gamma}=0.05,0.1,0.2,0.3,0.5,1$. The dashed line shows the distribution without noise. (b) $D/\dot{\gamma}=0.1$ and $e=1.5,2,3,5$. }
\label{fig:2Ddists}
\end{figure}
We obtain a numerical solution by taking the first $n$ terms in the expansion, Eq. (\ref{eq:harexp}), giving $2n$ equations.  We then set $a_{n+1}=0$ and $b_{n+1}=0$ in the last two equations with $k=n$. We have a linear system that can be solved in the unknowns $a_i,b_i,\;i=1,n$. 
Convergence of the series is rapid, but more terms are required as the elongation increases.  
With $e=8$ in the noiseless case, for $n=8$ spurious oscillations are present, but for $n=24$ the series solution is indistinguishable from the exact solution, Eq. (\ref{eq:psi0}). Some examples of the distribution are shown in Fig. \ref{fig:2Ddists}.

%fig12
 \begin{figure}[htbp]
 \begin{center}
  \includegraphics[width=8.5cm]{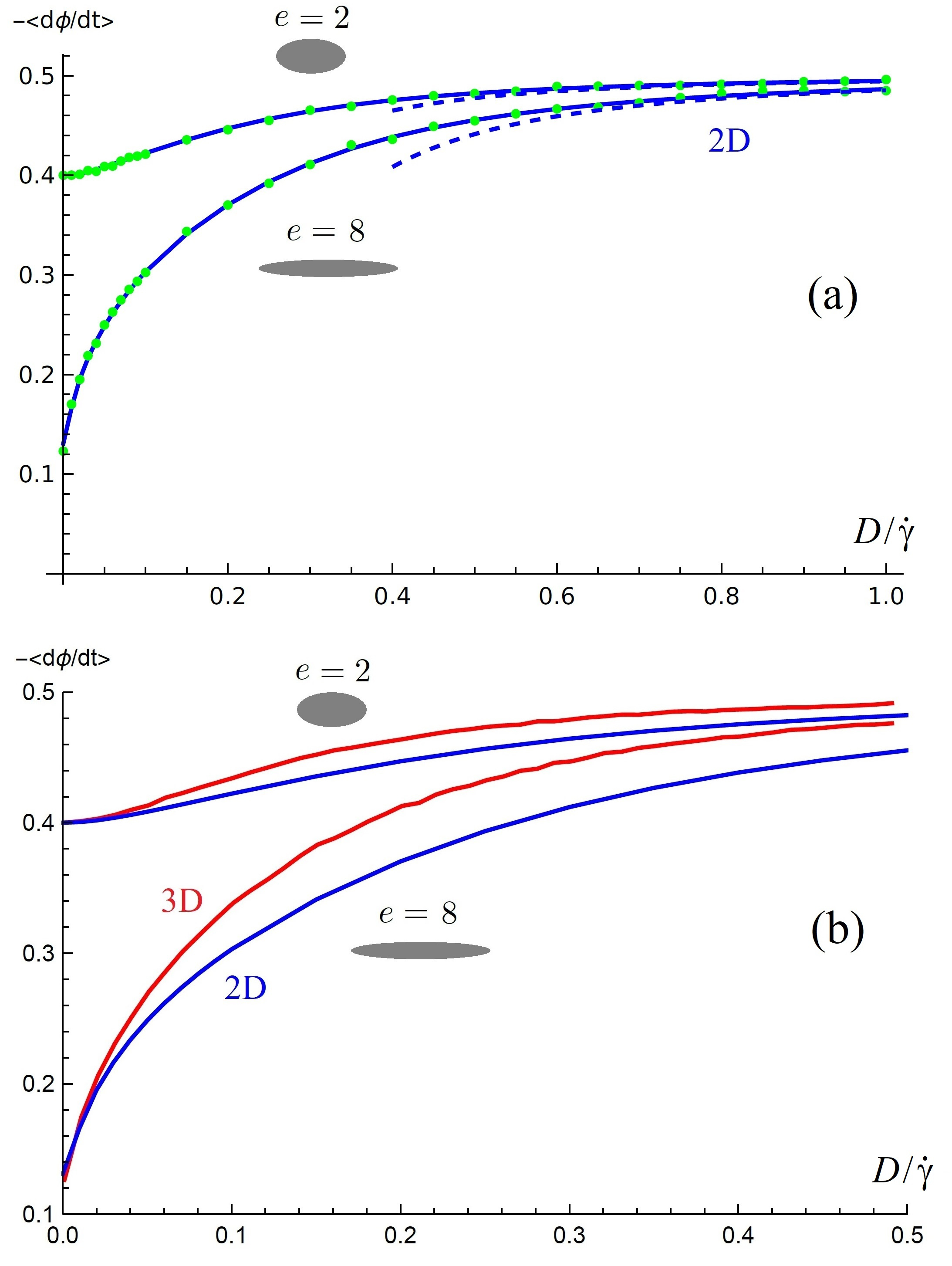}
 \end{center}
   \caption{(a) Mean angular velocity for particle elongations $e=2,8$ in two dimensions (2D) from the Fokker-Planck approach (solid lines) and Langevin simulations (points). The limiting values for small $D$ are 
   correctly given by Eq. (\ref{eq:fdot2Dzero}). The dashed curves show the perturbation theory to second order.
   (b) Comparison of mean angular velocity in 2D and 3D for $e=2,8$.}
   \label{fig:omegabar}
 \end{figure}

\subsection{Perturbation Theory}

We can apply singular perturbation theory to the Fokker-Planck equation (\ref{eq:TE}) to obtain the behavior at small noise. Taking $\epsilon=D/\dot{\gamma}$ as a small parameter:
\begin{equation}
\psi(\phi,\epsilon)=\psi_0(\phi)+\epsilon\psi_1(\phi)+\epsilon^2\psi_2(\phi)+\cdots,
\end{equation}
where $\psi_0(\phi)$ is the solution to the unperturbed problem.
Substituting in Eq. (\ref{eq:TE1}) we obtain
\begin{equation}
f(\phi)(\psi_0+\epsilon\psi_1+\cdots)=\epsilon\frac{d\psi_0}{d\phi}+\epsilon^2\frac{d\psi_1}{d\phi}+\cdots+c.
\end{equation}
Equating terms with the same power of $\epsilon$ we obtain
\begin{equation}
f(\phi)\psi_0(\phi)=c,\;\;f(\phi)\psi_1(\phi)=\frac{d\psi_0}{d\phi},\;\;f(\phi)\psi_2(\phi)=\frac{d\psi_1}{d\phi},\cdots
\end{equation}
Explicitly, at first order we find
\begin{equation}
\psi_1(\phi)=-\frac{e \left(e^4-1\right) \sin 2 \phi }{\pi  \left(\left(e^2-1\right) \sin
   ^2\phi +1\right)^3}.
\end{equation}

% Some examples of these functions are shown in Fig. \ref{fig:psik} for $e=1.5$. 
We note that  $\int_{-\pi/2}^{\pi/2}d\phi\;\psi_0(\phi) = 1$ and $\int_{-\pi/2}^{\pi/2}d\phi\;\psi_1(\phi) = 0$, but $\int_{-\pi/2}^{\pi/2}d\phi\;\psi_2(\phi) \ne 0$. Thus perturbation theory to first order in $\epsilon$, $\psi(\phi,\epsilon)=\psi_0(\phi)+\epsilon\psi_1(\phi)$, gives directly a normalized distribution. Adding the second order term, $\psi_2(\phi)$ results in a non-normalized distribution. One could, of course, renormalize this result. Also note that, even for a modest elongation of $e=1.5$ the amplitude of $\psi_k(\phi)$ increases rapidly with $k$. So we expect to obtain good results only for small $\epsilon$.

Similarly, we can also use perturbation theory to obtain the behavior at large noise by taking $Pe$ as the expansion parameter:
\begin{equation}
\psi(\phi,Pe)=\psi_0+Pe\psi_1(\phi)+Pe^2\psi_2(\phi)+\cdots
\end{equation}
where $\psi_0=1/\pi$ is the (uniform) distribution in the limit of large noise. Following the same procedure as above we obtain
\begin{eqnarray}
\psi_1(\phi)&=&\frac{\beta}{4\pi}\sin(2\phi)\\
\psi_2(\phi)&=&\frac{\beta}{32\pi}\cos(2\phi)(2-\beta\cos(2\phi)).
\end{eqnarray}
From these results we calculate
\begin{equation}
\langle\cos(2\phi)\rangle=\frac{\beta}{32}Pe^2+O(Pe^3).
\label{eq:PertPe2D}
\end{equation}

The corresponding results for the 3D system are presented in Appendices C and D.

\subsection{Mean angular velocity}
To find the mean angular velocity we take the average of Eq. (\ref{eq:FP2D})
\begin{equation}
\left<\frac{d\phi}{dt}\right>= -\frac{\dot{\gamma}}{2}(1-\beta\langle\cos2\phi\rangle) = -\frac{\dot{\gamma}}{2}(1-\frac{\pi}{2}\beta a_1).
\label{eq:omegabar2D}
\end{equation}
For $\epsilon=0$, $a_1=\frac{2}{\pi}\frac{e-1}{e+1}$ giving 

\begin{equation}
\left<\frac{d\phi}{dt}\right>_{\epsilon=0} = -\dot{\gamma}\frac{e}{e^2+1},
\label{eq:fdot2Dzero}
\end{equation}
which is the same as the 3D system, Eq. (\ref{eq:fdot3Dzero}).
In the limit of large noise we find
\begin{equation}
\left<\frac{d\phi}{dt}\right>_{\epsilon=\infty} = -\frac{\dot{\gamma}}{2}.
\end{equation}
The numerical results shown in Fig. \ref{fig:omegabar}(a) vary smoothly between these two limits. We also confirm that a perturbation theory estimate, obtained by substituting Eq. (\ref{eq:PertPe2D}) in Eq. (\ref{eq:omegabar2D}), describes the behavior at large noise. 

It is instructive to compare the two-dimensional system with the three-dimensional one, Fig. \ref{fig:omegabar}(b). The limiting behavior for small and large values of $D/\dot{\gamma}$ is the same, but for intermediate values, the mean angular velocity is higher in 3D.

%combined figure for angular velocity

%separate figures for angular velocity:
% \begin{figure}
% \begin{center}
%  \includegraphics[width=10cm]{Figures/2024-02-22-fig11-OmegaBar2Da-v1.jpg}
%   \includegraphics[width=10cm]{Figures/OmegaBar2Da.pdf}
% \end{center}
%   \caption{Mean angular velocity for $e=2,8$ in two dimensions from theory (solid lines) and Langevin simulations (points). The limiting values for small $D$ are    correctly given by Eq. (\ref{eq:fdot2Dzero}). The dashed curves show the perturbation theory to second order.}
   %\label{fig:omegabar2D}
% \end{figure}

%\begin{figure}
% \begin{center}
%  \includegraphics[width=10cm]{Figures/2024-02-22-fig12-OmegaBar2Dvs3Da-v1.jpg}
  %\includegraphics[width=10cm]{Figures/OmegaBar2Dvs3Da.pdf}
% \end{center}
%   \caption{Mean angular velocity for $e=2,8$ in 2D (blue) and 3D (red).}
%   \label{fig:omegabar2Dvs3D}
% \end{figure}

\subsection{Orientational Order Matrix}
 
 To quantify the ordering we can evaluate the orientational order matrix
\begin{equation}
Q=\left[
\begin{array}{cc}
 \langle\cos2\theta\rangle & \langle\sin2\theta\rangle  \\
 \langle\sin2\theta\rangle & -\langle\cos2\theta\rangle \\
\end{array}
\right]
\end{equation}
The matrix has two eigenvalues, $\pm S$, where $S$ is the nematic order parameter, and the corresponding eigenvectors are
\begin{equation}
\left(\begin{array}{c}
\cos\theta_p\\
\sin\theta_p\\
\end{array}
\right)
{\rm and}
\left(\begin{array}{c}
-\sin\theta_p\\
\cos\theta_p\\
\end{array}
\right)
\end{equation}
Due to asymmetry, the maximum of the distribution does not necessarily occur at $\theta=\theta_p$. The angle at which the distribution displays a maximum,  $\theta_{\rm max}$, and the eigenvector orientation, $\theta_p$, are equal in the limits of large and small noise. At intermediate noise,  $\theta_p < \theta_{\rm max}$. 
% Note that $\theta_p$ corresponds to $\theta_2$ 
% of the Marschall et al. paper. 

\begin{figure}
 \begin{center}
  \includegraphics[width=8.5cm]{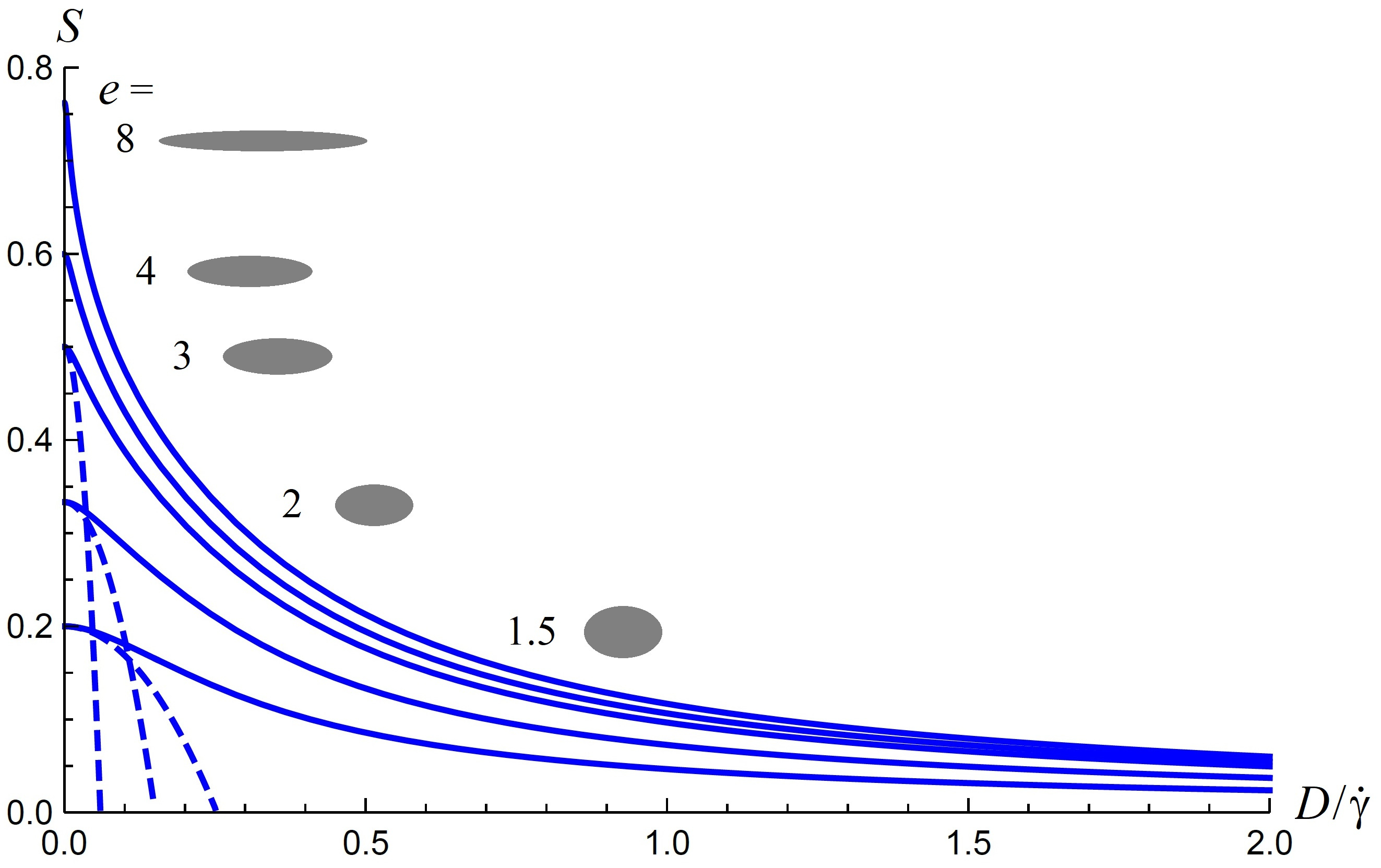}
 \end{center}
   \caption{Nematic order parameter $S$ as a function of the noise for $e=1.5, 2, 3, 4, 8$, bottom-to-top. The dashed lines show perturbation theory to second order, Eq. (\ref{eq:OPPert}).}
   \label{fig:orderparameter}
 \end{figure}

When $\psi(\theta)$ is expressed using the circular harmonic expansion, Eq. (\ref{eq:harexp}), the ordering matrix is
\begin{equation}
Q=\frac{\pi}{2}\left[
\begin{array}{cc}
 a_1 & b_1  \\
 b_1 & -a_1 \\
\end{array}
\right]
\end{equation}
Applying perturbation theory to second order results in 
\begin{equation}
S=\frac{e-1}{e+1}-c(e)\epsilon^2+O(\epsilon^3)
\label{eq:OPPert}
\end{equation}
with $c(e)=(1+e^2)^2(5e^8+2e^6-8e^5+8e^3-2e^2-5)/32e^6$. This, however, provides a good description only for weakly elongated particles at small values of $\epsilon$, see Fig.~\ref{fig:orderparameter}.

%Fig13
The orientation of the director, $\theta_p$,  can be computed from
\begin{equation}
\tan2\theta_p=\frac{\langle\sin2\theta\rangle}{\langle\cos2\theta\rangle}=\frac{b_1}{a_1}
\end{equation}
To obtain the Taylor series expansion to order $\epsilon$ we proceed as above with the result:
\begin{equation}
\tan2\theta_p=\frac{(1+e)^2(1+e^2)}{2e^2}\epsilon + O(\epsilon^2)
\end{equation}

\begin{figure}
\begin{center}
\includegraphics[width=8.5cm]{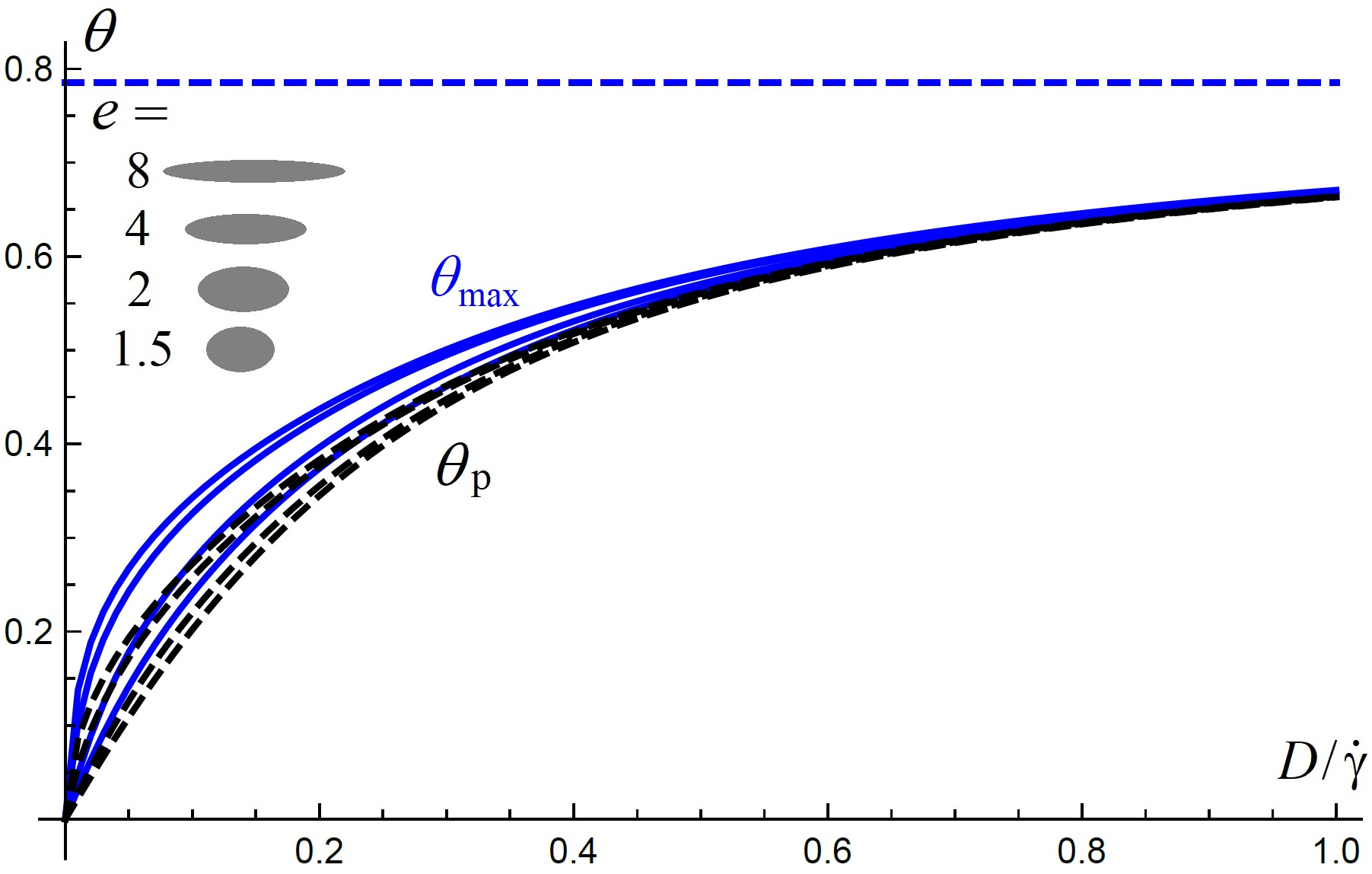}
% \subfloat[]{\includegraphics[width = 7cm]{Figures/thetamax5.pdf}} 
% \subfloat[]{\includegraphics[width = 7cm]{Figures/thetap5.pdf}}
\end{center}
%\includegraphics[width = \columnwidth]{Figures/thetamax5.pdf}\\ 
%\includegraphics[width = \columnwidth]{Figures/thetap5.pdf}
% \subfloat[$\beta=0.5$]{\includegraphics[width = 3in]{Figures/NOPbeta05.pdf}}
% \subfloat[$\beta=0.5$]{\includegraphics[width = 3in]{Figures/BOPbeta05.pdf}} 
\caption{ Angle for which the distribution displays a maximum $\theta_{\rm max}$, (blue, solid) and $\theta_p$ (black, dashed) as a function of the noise for $e=8, 4, 2, 1.5$ top-to-bottom calculated from the exact solution. 
   For large noise, both approach $\pi/4$ (dashed line).}
\label{fig:tmaxtp}
\end{figure}

Fig. \ref{fig:tmaxtp} shows that $\theta_p$ increases monotonically with $\epsilon=D/\dot{\gamma}$ and approaches an asymptotic value of $\pi/4$ in the limit of large noise. While $\theta_p$ shows some variation with elongation at low noise levels, it is insensitive to elongation at higher levels of noise.

%fig14
%\lipsum[1-2]

\section{Conclusion}

Jeffery orbits with noise can describe a variety of systems with anisometric particles under shear. In this contribution we extended the analytical result of Doi and Edwards for infinitely thin needles to particles with arbitrary shape factor, $\beta$. We examined the orientation order matrix as a function of the noise (or inverse P\'eclet number) and the particle elongation. The solutions of the Fokker-Planck equation agree with numerical simulations of the Langevin equation. Nematic ordering increases monotonically with increasing P\'eclet number at fixed shape and with increasing orientation at fixed P\'eclet number. By contrast, the biaxiality obtained for an elongated particle ($\beta>6$) exhibits a maximum for $Pe\approx 1.2$.  We also examined the behavior of the strictly two-dimensional system. Comparing the 2D and 3D data we find larger average rotation speed of the particle in a three-dimensional system for the same noise amplitude.

Systems of sheared inelastic particles under shear flow \cite{borzsonyiPRL2012} show alignment. These are typically dense systems with significant particle-particle interactions. In two-dimensions, it appears that, at least quantitatively, the orientational distribution of a system of inelastic hard dumbbells \cite{reddy2009orientational} resembles that of the Jeffery orbit with constant noise. It would be interesting to see if the same applies to the three dimensional systems considered in e.g. \cite{borzsonyiPRL2012,yuan1997non}. If there are significant differences can these systems be modeled using Jeffery orbits with an orientation dependent noise?
\vspace{3cm}
\begin{acknowledgments}

We acknowledge support from CNRS PICS Grant No. 08187 and the Hungarian Academy of Sciences Grant No. NKM-2018-5.
JT thanks V. Kumaran and Hartmut L\"owen for helpful discussions.

\end{acknowledgments}

\begin{widetext}

\appendix
%\section{Appendices}

\section{From Stratonovich to Itô expression of the Langevin equation}\label{appendixStrato}

The usual Stratonovich expression of the rotational diffusion (without any deterministic part)
\begin{equation}
\begin{array}{cc}
\textrm{(Strato) } & d\textbf{p} = \textbf{B}\left(\textbf{p}\right)\circ\boldsymbol{\xi}\,dt\end{array}
\end{equation}
%\[
%\begin{array}{cc}
%\textrm{(Strato) } & d\textbf{p}=\sqrt{2D} \, \textbf{p}\underset{Strato}{\otimes}\boldsymbol{\xi}\,dt=\textbf{B}\left(\textbf{p}\right)\cdot\boldsymbol{\xi}\,dt\end{array}
%\]
with 
\begin{equation}
\textbf{B}\left(\textbf{p}\right)=\sqrt{2D}\left(\begin{array}{ccc}
0 & -p_{z} & p_{y}\\
p_{z} & 0 & -p_{x}\\
-p_{y} & p_{x} & 0
\end{array}\right)
\end{equation}
can be turned into an Itô expression with the help of Itô's formula
\begin{equation}
\begin{array}{cc}
\textrm{(Itô) } & \left(d\textbf{p}\right)_{i}=\sqrt{2D}\left(\textbf{p}\times\boldsymbol{\xi}\right)_{i}\,dt+\frac{1}{2}\sum_{j,k}B_{kj}\partial_{k}B_{ij}\,dt\end{array}
\end{equation}
where the last term (drift induced by the multiplicative noise $\textbf{B}\left(\textbf{p}\right)\boldsymbol{\xi}\,dt$) can also be written as $-2Dp_{i}\,dt$, giving:
\begin{equation}
\begin{array}{cc}
\textrm{(Itô) } & \frac{d\textbf{p}}{dt}=\sqrt{2D} \, \textbf{p}\times\boldsymbol{\xi}-2D\textbf{p}\end{array}
\end{equation}

Adding the deterministic part (Jeffery orbits equation) does not modify the expression of the drift term $-2D\textbf{p}$
\begin{equation}
\begin{array}{cc}
\textrm{(Strato) } & \frac{d\textbf{p}}{dt}=\left(\textbf{I}-\textbf{p}\otimes\textbf{p}\right)\left(\beta \textbf{E}+\textbf{W}\right)\textbf{p}+\textbf{B}\left(\textbf{p}\right)\circ\boldsymbol{\xi}\,dt\end{array}
\end{equation}
\begin{equation}
\begin{array}{cc}
\textrm{(Itô) } & \frac{d\textbf{p}}{dt}=\left(\textbf{I}-\textbf{p}\otimes\textbf{p}\right)\left(\beta \textbf{E}+\textbf{W}\right)\textbf{p}+\sqrt{2D}\textbf{p}\times\boldsymbol{\xi}-2D\textbf{p}\end{array}
\end{equation}
since the latter arises only from the multiplicative noise term, independently from the deterministic part.

The passage from Cartesian $\left(p_{x},p_{y},p_{y}\right)$ to spherical $\left(p_{r},p_{\theta},p_{\phi}\right)$ coordinates is also readily obtained with the help of the Itô's formula for a change of variables (see \cite{Gardiner2004book}):
\begin{equation}\label{eq:ItoFormula}
\frac{df\left(\textbf{p}\right)}{dt}=\boldsymbol{\nabla} f\cdot\textbf{A}+\boldsymbol{\nabla} f\cdot\textbf{B}\boldsymbol{\xi}+\frac{1}{2}\sum_{i,j}\left(\textbf{B}\textbf{B}^{\top}\right)_{ij}\partial_{i}\partial_{j}f\left(\textbf{p}\right)
\end{equation}
for the general case where $\frac{d\textbf{p}}{dt}=\textbf{A}\left(\textbf{p}\right)+\textbf{B}\left(\textbf{p}\right)\boldsymbol{\xi}$.

%(\ref{eq:ItoFormula})

One gets (without the deterministic part)
\begin{align}
\frac{d\theta}{dt} & =\frac{d}{dt}\left(\arccos\left(p_{z}\right)\right)=D\cot\left(\theta\right)+\sqrt{2D}\left[\sin\left(\phi\right)\xi_{x}-\cos\left(\phi\right)\xi_{y}\right]\\
\frac{d\phi}{dt} & =\frac{d}{dt}\left(\arctan\left(p_{y}/p_{x}\right)\right)=\frac{\sqrt{2D}}{\sin\left(\theta\right)}\left[\cos\left(\theta\right)\left(\cos\left(\phi\right)\xi_{x}+\sin\left(\phi\right)\xi_{y}\right)-\sin\left(\theta\right)\xi_{z}\right]
\end{align}

We can define three new Brownian motions 
\begin{align}
\xi_{1} & =\sin\left(\phi\right)\xi_{x}-\cos\left(\phi\right)\xi_{y}\\
\xi_{2} & =\cos\left(\theta\right)\left(\cos\left(\phi\right)\xi_{x}+\sin\left(\phi\right)\xi_{y}\right)-\sin\left(\theta\right)\xi_{z}\\
\xi_{3} & =\sin\left(\theta\right)\left(\cos\left(\phi\right)\xi_{x}+\sin\left(\phi\right)\xi_{y}\right)+\cos\left(\theta\right)\xi_{z}
\end{align}
as linear combinations of the independent Brownian motions $\left(\xi_{x},\xi_{y},\xi_{z}\right)$ through an orthogonal matrix $\tilde{\textbf{S}}$
\begin{equation}
\tilde{\textbf{S}}=\left(\begin{array}{ccc}
\sin\phi & -\cos\phi & 0\\
\cos\theta\cos\phi & \cos\theta\sin\phi & -\sin\theta\\
\sin\theta\cos\phi & \sin\theta\sin\phi & \cos\theta
\end{array}\right)
\end{equation}
Then, it is easy to check that they are independent (see \cite{Gardiner2004book}) and orthonormal ($\xi_{i} \cdot\xi_{j}=\delta_{ij}$ in the $\left\{ \xi_{x},\xi_{y},\xi_{z}\right\} $ basis).

%where one can easily check that the two new Brownian motions $\xi_{1}=\sin\left(\phi\right)\xi_{x}-\cos\left(\phi\right)\xi_{y}$ and $\xi_{2}=\cos\left(\theta\right)\left(\cos\left(\phi\right)\xi_{x}+\sin\left(\phi\right)\xi_{y}\right)-\sin\left(\theta\right)\xi_{z}$ are independent and orthogonal.

One finally obtains the expressions
\begin{align}
\frac{d\theta}{dt} & = D\cot\theta+\sqrt{2D} \, \xi_{1}\\
\frac{d\phi}{dt} & = \frac{\sqrt{2D}}{\sin\theta} \, \xi_{2}
\end{align}
to which the deterministic Jeffery's terms must be added to get Eq. (15-18).
%(\ref{eq:thetaphieqShearxy}-\ref{eq:thetaphieqShearxz})

The conservation of the norm of the unit vector $\textbf{p}$ can also be readily seen using Itô's formula for $f\left(\textbf{p}\right)=\textbf{p} \cdot \textbf{p} = p^2$. In this case, the last term of Eq. (\ref{eq:ItoFormula}) becomes
\begin{equation}
\frac{1}{2}\sum_{i,j}\left(\textbf{B}\textbf{B}^{\top}\right)_{ij}\partial_{i}\partial_{j}f\left(\textbf{p}\right)=4Dp^2
\end{equation}
compensating exactly the drift induced term $\boldsymbol{\nabla} f\cdot\textbf{A}$ in Eq. (\ref{eq:ItoFormula}), while the $\boldsymbol{\nabla} f\cdot\textbf{B}\boldsymbol{\xi}$ term vanishes since 
\begin{equation}
\boldsymbol{\nabla} f\cdot\textbf{B}\boldsymbol{\xi}=-2\sqrt{2D}\textbf{p}\cdot\left(\textbf{p}\times\boldsymbol{\xi}\right)=0
\end{equation}
Finally
\begin{equation}
\frac{d\left(p^2\right)}{dt}=2\textbf{p}\cdot\left(\textbf{I}-\textbf{p}\otimes\textbf{p}\right)\left(\beta \textbf{E}+\textbf{W}\right)\textbf{p}=2\left(1-p^2\right)\textbf{p}\cdot\left(\beta \textbf{E}+\textbf{W}\right)\textbf{p}
\end{equation}
showing that $\frac{d\left(p^2\right)}{dt}=0$ at any time if $p^2=1$ at the initial time.

\section{Computation of the $\left(lm|\hat{\Lambda}|l'm'\right)$}\label{appendixMatrix}

In this appendix we detail how to compute the $\left(lm|\hat{\Lambda}|l'm'\right)$ matrix elements necessary for the evaluation of the $b_{lm}$ coefficients of the distribution function $\psi\left(\theta,\phi\right)$ appearing in Eq. (\ref{eq:psiFP}) and evaluated in Eq. (\ref{eq:blm}).

As shown in the main text, the operator $\hat{\Lambda}$ can be written as $\hat{\Lambda}=\beta\hat{\Lambda}_{1}+\frac{1-\beta}{2}i\hat{L}_{y}$ for an elongated particle of shape factor $\beta$. The operator $\hat{\Lambda}_{1}$ introduced by Doi and Edwards in \cite{doi1978dynamics} depends on the usual spherical harmonics $Y_{l}^{m}$ and angular momentum operators $\hat{L}_{z}$ and $\hat{L}_{y}$.

For the contribution from $\hat{L}_{z}$ and $i\hat{L}_{y}$, we use
\begin{equation}
\langle l,m|\hat{L}_{z}|l',m'\rangle=m\,\delta_{l,l'}\,\delta_{m,m'}
\end{equation}
%Let us underline that the basis functions |l,m) are “not” eigenvectors of the operator \hat{L}_{z}.
and
\begin{equation}
\langle l,m|i\hat{L}_{y}|l',m'\rangle=\frac{1}{2}\sqrt{l(l+1)-m(m-1)}\,\delta_{l,l'}\,\delta_{m-1,m'}-\frac{1}{2}\sqrt{l(l+1)-m(m+1)}\,\delta_{l,l'}\,\delta_{m+1,m'}
\end{equation}
leading to
\begin{equation}
(l,m|i\hat{L}_{y}|l',m')=\frac{g_{m}}{2}\left[\sqrt{l(l+1)-m(m-1)}g_{m-1}\,\delta_{m-1,m'}-\sqrt{l(l+1)-m(m+1)}g_{m+1}\,\delta_{m+1,m'}\right]\,\delta_{l,l'}
\end{equation}
where $g_{0}=\sqrt{2}$, $g_{m}=1$ if $m>0$, and $g_{m}=0$ if $m<0$.

For the terms implying the spherical harmonics in $\hat{\Lambda}_{1}$, we use the relation
\begin{equation}
\langle l,m|Y_{p}^{q}|l',m'\rangle=\left(-1\right)^{m}\sqrt{\frac{\left(2l+1\right)\left(2p+1\right)\left(2l'+1\right)}{4\pi}}\left(\begin{array}{ccc}
l & p & l'\\
0 & 0 & 0
\end{array}\right)\left(\begin{array}{ccc}
l & p & l'\\
-m & q & m'
\end{array}\right)
\end{equation}
in terms of the Wigner 3-j symbols 
\begin{equation}
\left(\begin{array}{ccc}
j_{1} & j_{2} & j_{3}\\
m_{1} & m_{2} & m_{3}
\end{array}\right)=\frac{\left(-1\right)^{j_{1}-j_{2}-m_{3}}}{\sqrt{2j_{3}+1}}\langle j_{1},m_{1};j_{2},m_{2}|j_{3},m_{3}\rangle
\end{equation}
which are different from $0$ only if $\left|j_{1}-j_{2}\right|\leq j_{3}\leq j_{1}+j_{2}$,
$m_{3}=-m_{1}-m_{2}$ and $-j_{k}\leq m_{k}\leq j_{k}$ for $k=1,2,3$.

As noted in \cite{doi1978dynamics}, only a few $(l,m|\hat{\Lambda}_{1}|l',m')$ terms are different from zero in the summation of Eq. (\ref{eq:blm}). These are
\begin{equation}
(l,m|\hat{\Lambda}_{1}|l,m-1)=g_{m-1}G_{1}\left(l,m-1\right)
\end{equation}

\begin{equation}
(l,m|\hat{\Lambda}_{1}|l,m+1)=-g_{m}G_{1}\left(l,-m-1\right)
\end{equation}

\begin{equation}
(l,m|\hat{\Lambda}_{1}|l+2,m-1)=g_{m-1}G_{2}\left(l,m-1\right)
\end{equation}

\begin{equation}
(l,m|\hat{\Lambda}_{1}|l+2,m+1)=-g_{m}G_{2}\left(l,-m-1\right)
\end{equation}

\begin{equation}
(l,m|\hat{\Lambda}_{1}|l-2,m-1)=g_{m-1}G_{3}\left(l-2,m-1\right)
\end{equation}

\begin{equation}
(l,m|\hat{\Lambda}_{1}|l-2,m+1)=-g_{m}G_{3}\left(l-2,-m-1\right),
\end{equation}
where, as previously stated
\begin{equation}
g_{m}=\left\{ \begin{array}{ccc}
\sqrt{2} &  & \textrm{ if }m=0\\
1 &  & \textrm{ if }m>0\\
0 &  & \textrm{ if }m<0
\end{array}\right.
\end{equation}
and
\begin{equation}
G_{1}\left(l,m\right)=\frac{2l^{2}+2l+3m}{2\left(2l+1\right)\left(2l+3\right)}\left[\left(l-m\right)\left(l+m+1\right)\right]^{1/2}
\end{equation}

\begin{equation}
G_{2}\left(l,m\right)=\frac{l}{2\left(2l+3\right)}\left[\frac{\left(l+m+2\right)\left(l-m\right)\left(l-m+1\right)\left(l-m+2\right)}{\left(2l+1\right)\left(2l+5\right)}\right]^{1/2}
\end{equation}

\begin{equation}
G_{3}\left(l,m\right)=\frac{l+3}{2\left(2l+3\right)}\left[\frac{\left(l-m+1\right)\left(l+m+1\right)\left(l+m+2\right)\left(l+m+3\right)}{\left(2l+1\right)\left(2l+5\right)}\right]^{1/2}
\end{equation}
following the notation of Doi and Edwards in \cite{doi1978dynamics}.

\section{Perturbation theory in 3D using $Pe$ as the expansion parameter}
\label{appendixPertPe}

Here we develop a perturbation theory using the P\'eclet number $Pe=\dot{\gamma}/D$ as the expansion parameter. The Fokker-Planck equation may be written as 
\begin{equation}
Pe \, \hat{\Lambda} \psi=\nabla^2\psi
\end{equation}
Let us expand the orientational distribution function as a power series in $Pe$:
\begin{equation}
\psi=\psi_0+Pe \, \psi_1+Pe^2 \, \psi_2+\cdots
\end{equation}
Substituting in the FP equation and equating powers of $Pe$ we obtain at zeroth, first  and second order 
\begin{eqnarray}
\nabla^2\psi_0&=&0\nonumber\\
\hat{\Lambda} \psi_0&=&\nabla^2\psi_1\nonumber\\
\hat{\Lambda} \psi_1&=&\nabla^2\psi_2
\end{eqnarray}
Solving we find
\begin{eqnarray}
\psi_0&=&\frac{1}{4\pi}\\
\psi_1&=&-\frac{\beta}{4\pi}\sqrt{\frac{\pi}{15}}|21)\\
\psi_2&=&-\frac{(7-\beta)\beta}{336\sqrt{5\pi}}|20)+\frac{(7+3\beta)\beta}{336\sqrt{15\pi}}|22)
-\frac{\beta^2}{420\sqrt{\pi}}|40)+\frac{\beta^2}{168\sqrt{5\pi}}|42)
\end{eqnarray}
which gives the scalar order parameter (defined in section D)
\begin{equation}
s=\frac{1}{15}\beta Pe+O(Pe^2)
\label{eq:OPPTbeta}
\end{equation}
%Note that for small $Pe$ (or large $D/\dot{\gamma}$) Eq. (\ref{eq:OPPTbeta}) reduces to this expression.
Similarly we find
\begin{equation}
\langle\cos(2\phi)\rangle=\frac{(5+3\beta)\beta}{720}Pe^2+O(Pe^3)
\end{equation}

\section{Perturbation theory in 3D using $\beta$ as the expansion parameter}
\label{appendixPertBeta}

We know that when $\beta=0$, corresponding to a sphere, the orientational distribution is uniform: $\psi(\theta,\phi)=1/(4\pi)$. This prompts us to  try a perturbation approach. Let 
\begin{equation}
\psi(\beta,\theta,\phi)=\psi_0(\theta,\phi)+\beta\psi_1(\theta,\phi)+\beta^2\psi_2(\theta,\phi)+\cdots
\end{equation}
Substituting in the Fokker-Planck equation gives 
\begin{equation}
\left[\beta\hat{\Lambda}_1+\frac{1-\beta}{2}i\hat{L}_y\right]\psi(\beta,\theta,\phi)=\epsilon\nabla^2\psi(\beta,\theta,\phi)
\end{equation}
where $\epsilon=D/\dot{\gamma}$.
Expanding in $\beta$ and equating the coefficients of equal powers gives
\begin{eqnarray}
\frac{1}{2}i\hat{L}_y\psi_0=\epsilon\nabla^2\psi_0\\
\Lambda_1\psi_0+\frac{1}{2}i\hat{L}_y(\psi_1-\psi_0)=\epsilon\nabla^2\psi_1
\end{eqnarray}
Clearly $\psi_0=1/4\pi$ and the second equation becomes
\begin{equation}
3\sqrt{\frac{4\pi}{15}}\frac{S_2^1}{4\pi}+\frac{1}{2}i\hat{L}_y\psi_1=\epsilon\nabla^2\psi_1
\end{equation}
%\newpage
\nocite{*} 
Let us assume
\begin{equation}
\psi_1=b_{20}|20)+b_{21}|21)+b_{22}|22)
\end{equation}
Substituting in the above equation and using the orthonormality of the functions $|lm)$ we find 
\begin{eqnarray}
3\sqrt{\frac{4\pi}{15}}\frac{1}{4\pi}+\frac{\sqrt{3}}{2}b_{20}-\frac{1}{2}b_{22}=-6\epsilon b_{21}\\
-\frac{\sqrt{3}}{2}b_{21}=-6\epsilon b_{20}\\
\frac{1}{2}b_{21}=-6\epsilon b_{22}
\end{eqnarray}
Solving these equations  and using the coefficients to evaluate the order matrix we find
\begin{equation}
\mathbf{Q}=
\begin{pmatrix}
\frac{\beta}{5 ( 1+36\epsilon^2 ) }& 0&\frac{6(\epsilon)\beta}{5( 1+36\epsilon^2 )} \\
0 &  0 & 0 \\
\frac{6(\epsilon)\beta}{5( 1+36\epsilon^2 )}  & 0&-\frac{\beta}{5( 1+36\epsilon^2 ) }
\end{pmatrix}
\end{equation}
The eigenvalues are
\begin{equation}
\left(\frac{\beta}{5\sqrt{36\epsilon^2+1}},0,\frac{-\beta}{5\sqrt{36\epsilon^2+1}}\right)
\end{equation}
and hence we find the scalar order parameters 
\begin{equation}
s=2r=\frac{2\beta}{5\sqrt{1+36\epsilon^2}}
\end{equation}
Note that for small $Pe$ (or large $D/\dot{\gamma}$) Eq. (\ref{eq:OPPTbeta}) reduces to this expression.

\section{The scalar order parameters $s$ and $r$ for Jeffery orbits with a uniformly distributed random initial orientation}
\label{appendixOPNoNoise}

For a given initial orientation specified by $\phi_0,\theta_0$ the corresponding values of the parameters $C$ and $\kappa$ are given by
\begin{eqnarray}
C^2&=&\tan^2(\theta_0)(e^2\sin^2\phi_0+\cos^2\phi_0)/e^2\\
\kappa&=&\arctan(e\tan(\phi_0))
\end{eqnarray}
We can calculate the order matrix Eq. (\ref{eq:QMpsi}) by replacing the average over the orientational probability distribution by a time average over one period (actually symmetry allows us to limit this to one quarter period). We then average $Q$ over a sample of random initial orientations and then finally calculate $s$ and $r$ using the procedure given in the main text. 

However, a more efficient method is to determine the distribution of $y=C^2$.
Assuming a uniform, random orientation we obtain the probability density function
\begin{equation}
f(y)=\frac{e}{\pi(1+e^2y)\sqrt{1+y}}E\left(-\sqrt{\frac{(e^2-1)y}{1+y}}\right),\;0\le y<\infty
\end{equation}
where $E(x)$ is the complete elliptic integral of the second kind.
By integrating over one quarter period of the Jeffery orbit we can calculate the order matrix. The non-zero elements are 
\begin{eqnarray}
\langle\cos^2\theta\rangle&=&\frac{1}{\sqrt{(1+y)(1+e^2y)}}\\
\langle\cos^2\theta\sin^2\phi\rangle&=&\frac{e^2y}{1+e^2y+\sqrt{(1+y)(1+e^2y)}}\\
\langle\sin^2\theta\sin^2\phi\rangle&=&\frac{y}{1+y+\sqrt{(1+y)(1+e^2y)}}
\end{eqnarray}
We now average these matrix elements over the distribution of $y=C^2$ by multiplying them by $f(y)$ and integrating over the domain. It does not seem possible to obtain analytical expressions for the integrals, but they can be easily evaluated numerically. The order matrix constructed with these elements is diagonal. Evaluating $s$ and $r$  we obtain the results shown in Fig. \ref{fig:OPSZero}. We note that nematic ordering increases more and more rapidly as $\beta$ approaches one, while biaxiality displays a maximum value of 0.121 for $\beta=0.876$ before falling rapidly to zero.
\end{widetext}

%\newpage

 \bibliography{Jeffery}
 %\newpage
 %\section{Appendix A}
 %\newpage
 %\includepdf[pages=-]{AppendixA.pdf}
 %\includepdf[pages=-]{AppendixB.pdf}
 
\end{document}